\documentclass[conference, 10pt]{IEEEtran}
\usepackage[font=small]{caption}
\usepackage{graphicx, balance}
\usepackage{subcaption}
\usepackage{linguex}
\usepackage[linguistics,edges]{forest}
\usepackage{enumitem}
\usepackage{booktabs}
\usepackage{multirow, makecell}
\usepackage[cmex10]{amsmath}
\usepackage{algorithmic}
\usepackage{url}
\usepackage{csquotes} 
\usepackage{arydshln} 
\usepackage{color, xcolor, soul}
\usepackage{amssymb}
\usepackage{makecell}
\usepackage{cite}
\usepackage{etoolbox}
\usepackage{threeparttable}
\usepackage{hyperref}
\usepackage{pifont}

\usepackage{array}
\usepackage{booktabs}
\usepackage{multirow}
\usepackage{tabularx}

\usepackage{rotating}

\newcolumntype{C}[1]{>{\centering\arraybackslash}p{#1}}

\newcommand*\halfcirc[1][1ex]{%
  \begin{tikzpicture}
  \draw[fill] (0,0)-- (90:#1) arc (90:270:#1) -- cycle ;
  \draw[thick] (0,0) circle (#1);
  \end{tikzpicture}}
\newcommand*\fullcirc[1][1ex]{\tikz\fill (0,0) circle (#1);} 

\title{Phishing Webpage Detection: Unveiling the Threat Landscape and Investigating Detection Techniques}

\author{\IEEEauthorblockN{Aditya Kulkarni\textsuperscript{*}, Vivek Balachandran\textsuperscript{\#}, and Tamal Das\textsuperscript{*}}
\IEEEauthorblockA{\textsuperscript{*}\emph{Indian Institute of Technology, Dharwad, India} \\
\textsuperscript{\#}\emph{Singapore Institute of Technology, Singapore} \\
\textsuperscript{*}\{aditya.kulkarni, tamal\}@iitdh.ac.in, \textsuperscript{\#}vivek.b@singaporetech.edu.sg}}

\begin{document}


\maketitle

\begin{abstract}
In the realm of cybersecurity, \textit{phishing} stands as a prevalent cyber attack, where attackers employ various tactics to deceive users into gathering their sensitive information, potentially leading to identity theft or financial gain. Researchers have been actively working on advancing phishing webpage detection approaches to detect new phishing URLs, bolstering user protection. Nonetheless, the ever-evolving strategies employed by attackers, aimed at circumventing existing detection approaches and tools, present an ongoing challenge to the research community.

This survey presents a systematic categorization of diverse phishing webpage detection approaches, encompassing URL-based, webpage content-based, and visual techniques. Through a comprehensive review of these approaches and an in-depth analysis of existing literature, our study underscores current research gaps in phishing webpage detection. Furthermore, we suggest potential solutions to address some of these gaps, contributing valuable insights to the ongoing efforts to combat phishing attacks.

\end{abstract}


\section{Introduction}  
\label{sec:Introduction}
The affordability of mobile devices and the increasing reach of Internet services has brought about $66\%$ of the world population online~\cite{ituStatistics}. Users increasingly rely on the Internet to perform their daily and professional work. Such activities involve information sharing in the form of text, image, audio, and video across social media websites, e-commerce/banking websites, emails, etc. Online payments are also at an all-time high, which requires sharing of email credentials, credit card details, etc. Such rising Internet dependency increases the vulnerability of its user's sensitive information. Cybercrimes like identity theft, ransomware attacks, data breaches, financial losses, and phishing attacks are accomplished using malware, adware, trojans, botnets, SQL injection, man-in-the-middle attacks, and denial of service. Such widespread cybercrimes result in heavy economic losses and identity theft~\cite{itgovernanceBiggestPhishing}.

Cybersecurity is the \emph{art of protecting networks, devices, and data from unauthorized access or criminal use and ensuring confidentiality, integrity, and availability of information}~\cite{cisaWhatCybersecurity}. Several government and corporate organizations regularly sensitize their employees on the best measures for online security. Among the various cyber threats that loom large today, phishing attacks have emerged as one of the most pervasive and insidious forms of cybercrime. Phishing is a \textit{criminal mechanism employing both social engineering and technical subterfuge to steal consumers' identity data and financial account credentials}~\cite{frauenstein2020susceptibility}. Phishing is the most pervasive cybercrime used for information-gathering. An attacker replicates a legitimate website, causing users to unwittingly disclose their sensitive information, often as a result of phishing emails, SMS messages, or social media platforms.

\begin{figure}[t]
    \centering
    \includegraphics[width=0.49\textwidth]{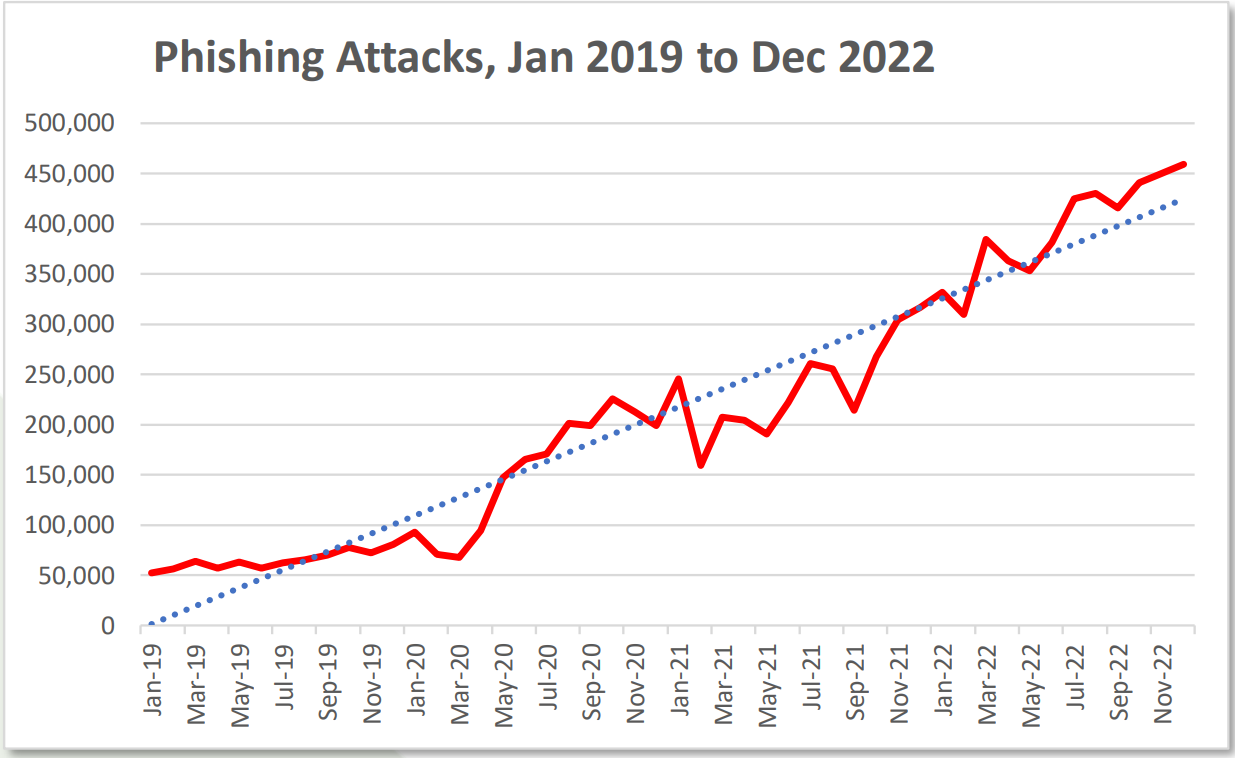}
    \caption{\centering Phishing Attacks from Jan $2019$ to Dec $2022$: APWG~\cite{APWG_REPORT_4_2022}}
    \label{fig:APWG_Report_4_Years}
\end{figure}

In the first quarter of $2018$, around $647,592$ phishing websites were created targeting legitimate financial websites~\cite{apwg_report_2018}. These phishing websites entice users into divulging their sensitive information while convincingly imitating legitimate websites. Moreover, the graph presented in Figure~\ref{fig:APWG_Report_4_Years} illustrates a significant rise in phishing attacks over a four-year period, starting from January $2019$ and continuing until December $2022$. The increase in attacks was notably steep, with an acceleration of more than $150\%$ in the number of attacks each year. This upward trend has raised significant concerns among cybersecurity experts and highlights the urgent need for robust defence mechanisms to counter this ever-evolving threat.

Detecting and mitigating phishing attacks require a multipronged approach that combines technology, user education, and continuous monitoring. In response to the escalating threat landscape, cybersecurity professionals have been developing and refining innovative detection approaches. Advanced email filtering systems leverage artificial intelligence and machine learning (ML) algorithms and browser plugins to identify suspicious emails and uniform resource locator (URL), flagging them for further inspection. Additionally, security awareness training for individuals and employees plays a pivotal role in reducing the success rate of phishing attacks. Educating users about the common red flags and best practices when handling emails can empower them to make informed decisions and avoid falling victim to phishing attempts. As cyber criminals continually adapt their tactics, cybersecurity experts must remain vigilant and adaptive, continuously upgrading their detection methods to stay one step ahead in this ongoing battle to secure the digital realm.

In the existing literature, various phishing webpage detection approaches have been proposed, ranging from traditional list-based methods~\cite{cao2008anti} to more advanced \textit{URL-based}~\cite{ma2009beyond} and \textit{webpage content analysis}~\cite{mao2017phishing} techniques. These advancements were prompted by the limitations observed in earlier detection approaches. For instance, the list-based approach lacked the ability to identify \textit{zero-day phishing attacks} (i.e., newly hosted phishing webpages), leading to the development of analysis techniques that focused on phishing URLs and webpage content. These improved methods aimed to effectively flag new phishing URLs and webpages based on their distinctive features.

In recent times, ML approaches have gained prominence, offering faster URL and content analysis capabilities and enabling the generation of precise decision rules for identifying new phishing URLs and webpages~\cite{alam2020phishing, jain2019machine}. Furthermore, significant strides have been made in visual analysis, particularly through deep learning (DL) approaches~\cite{abdelnabi2020visualphishnet}. These DL-based models excel at extracting features from webpage screenshots and logos, leading to more sophisticated and accurate webpage detection techniques. As researchers continuously refine and innovate in this field, the evolution of phishing webpage detection solutions continues to bolster cybersecurity efforts in combating this pervasive threat.

In this comprehensive literature review, we present an overview of phishing webpage detection approaches focusing on three main inputs: URLs, webpage content, and webpage screenshots. We delve into the datasets employed, the feature sets utilized, the feature selection algorithms applied, and the performance metrics adopted for existing phishing webpage detection solutions. Additionally, we highlight the uniqueness of our work concerning previous survey papers in the domain of phishing webpage detection solutions. We achieve this by identifying and addressing open issues prevalent in the current solutions and proposing potential methods and solutions to tackle these challenges effectively. Our study aims to contribute to the advancement of phishing webpage detection methods and foster a deeper understanding of the state-of-the-art techniques in this critical area of cybersecurity.


\subsection{Our Contribution}
\label{sec:Contribution}
In this paper, we provide the following contributions:
\begin{enumerate}
    \item We offer a comprehensive and structured review of current phishing webpage detection approaches by categorizing them based on their input sources, including URL analysis, webpage content examination, and screenshot-based approaches.
    \item We explore a diverse array of features, and dataset repositories for effectively distinguishing phishing attacks from legitimate, enhancing the understanding of dataset collection for selective effective features in developing efficient phishing webpage detection models.
    \item We identify and discuss various issues associated with existing phishing webpage detection approaches, presenting potential solutions for some of these challenges. We also highlight recent challenges emerging in the domain, for which only a limited number of solutions have been proposed. This helps in highlighting areas that require further research and development.
\end{enumerate}

\subsection{Organization of Paper}
\label{sec:Organization_of_Paper}
The remainder of this paper is as follows. Section~\ref{sec:Background} defines various aspects of phishing webpage detection solutions, including inputs for phishing, dataset repositories with their sample types, features, feature selection algorithms, ML and DL algorithms, and performance metrics. Section~\ref{sec:Methodology_Taxonomy_for_Phishing_Detection}, provides an overview of the methodologies employed in phishing webpage detection. In Section~\ref{sec:Discussion}, we present our critique on the topics covered in the previous sections, offering an overall understanding of the domain. Section~\ref{sec:Current_Hurdles_in_Phishing_Detection} provides challenges and the latest tactics on phishing webpage detection models with their overcoming solutions. Section~\ref{sec:Open_Issues} delves into open issues related to datasets, features, and approaches, along with their potential solutions. In Section~\ref{sec:Contrast_with_Existing_Surveys}, we analyze the existing surveys and contrast them with our work. Finally, we conclude our survey in Section~\ref{sec:Conclusion_and_Future_Work}.

\section{Background}
\label{sec:Background}
This section provides a concise overview of essential concepts for developing phishing webpage detection models capable of accurately classifying new samples into either the ``phishing'' or ``legitimate'' class. It begins by shedding light on the necessary inputs for phishing webpage detection approaches. Subsequently, the discussion revolves around the selection of existing repositories used to gather phishing and legitimate samples. The section thoroughly discusses key features that play a pivotal role in distinguishing phishing from legitimate samples. To enhance classifier training, feature selection algorithms are employed. Additionally, the section explores ML classifiers and DL algorithms that are essential components of effective phishing webpage detection systems. Towards the end, the significance of performance metrics is emphasized. These metrics provide valuable insights into how well a model can accurately classify new samples, ultimately contributing to the overall effectiveness of the phishing webpage detection system. By combining different detection approaches, feature selection techniques, and performance analysis, this section equips readers with the knowledge to develop robust and accurate ML-based phishing webpage detection solutions and understand subsequent sections.

\subsection{Inputs for Phishing Webpage Detection}
\label{sec:Inputs_for_Phishing_Detection}
Phishing attacks typically involve malicious links sent to target users through channels like emails, SMS, or social media platforms. These links resemble legitimate domain names, tricking users into clicking and redirecting them to phishing web pages. These deceptive pages often mimic the appearance of legitimate sites and use screenshots of the original content, overlaying input fields to collect sensitive user credentials.

To effectively combat phishing, a comprehensive analysis of the phishing URL and its landing web page is essential. Visual inspection includes examining the favicon, logo, and textual content for any discrepancies. Additionally, scrutinizing Javascript codes, CSS rules, and PHP codes in the landing page helps identify suspicious elements that may hint at phishing attempts.

Advanced phishing webpage detection solutions also consider webpage certificates and the presence of a padlock icon in the address bar. These elements add value to the analysis by providing insights into the ownership and security of the hosted webpage. By combining these techniques, users and organizations can better protect themselves against phishing threats and safeguard sensitive information from falling into the hands of attackers.

\subsubsection{URL}
\label{sec:URL}
\begin{figure}[h]
    \centering
    \includegraphics[width=0.49\textwidth, height=0.18\textwidth]{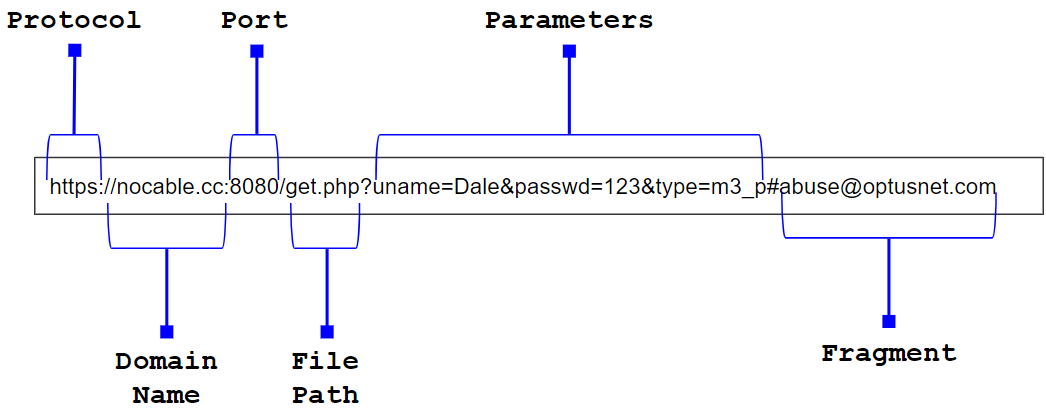}
    \caption{URL Structure}
    \label{fig:URL_Structure}
\end{figure}

A URL consists of distinct components that collectively define how to locate and access resources on the internet. These components are as follows:
\begin{enumerate}
    \item \textit{Protocol:} Serving as the first element, the protocol dictates the rules and method for resource retrieval from the server. Examples include HTTP and HTTPS for web pages, FTP for file transfers, etc.
    \item \textit{Domain Name:} Positioned after the protocol, the domain name identifies the specific host or server where the resource resides. It can be an organization's name, a website, or an IP address (IPv$4$ or IPv$6$).
    \item \textit{Port Number:} Optionally following the domain name and separated by a colon, the port number designates the communication endpoint on the server for the client to connect. If omitted, the default port associated with the protocol is assumed.
    \item \textit{File Path:} Representing the hierarchical directory structure, the path comes after the domain name and leads to the exact location of the resource on the server.
    \item \textit{Parameters:} Optionally placed after the path and initiated with a question mark ($?$), the parameter (query) contains key-value pairs, allowing the passing of parameters to the server for resource customization or specific data retrieval.
    \item \textit{Fragment:} Optionally indicated by a hash ($\#$) symbol and placed at the end, the fragment points to a specific section or anchor within the resource, often used in web pages to direct users to a particular part of the page.
\end{enumerate}
An example of a URL with all its components is shown in Figure~\ref{fig:URL_Structure}.

A URL with the HTTPS protocol indicates that the connection is secure with Secure Socket Layer (SSL) certification, currently utilizing Transport Layer Security (TLS). The presence of a padlock icon in the URL's address bar confirms the secure connection to the domain server. Certificates for webpages are issued by Certificate Authorities (CAs) and contain specific fields such as \textit{Common Name (CN)}, \textit{Organization (O)}, and \textit{Organizational Unit (OU)}. These details appear in two sections: \textit{Issued To} (say, domain owner) and \textit{Issued By} (say, issuer), along with information about the certificate's \textit{validity period} and \textit{fingerprints}.
In the context of URLs and certificate levels, a hosted web page invariably includes details of the certificate, which in turn comprises information about the issuer and domain owner. Each webpage falls into one of three categories of SSL certification: \textit{Domain Validated (DV)}, \textit{Organization Validated (OV)}, and \textit{Extended Validation (EV)}. The first level of validation is DV, granting the webpage owner control over the domain. The Certificate Authority (CA) verifies the webpage by sending an email to the owner for confirmation. The next level is OV, which enhances the validation process with additional layers. CAs at this level seek the business name, status, type, and physical address of the domain owner for validation. The highest level is EV, which involves a rigorous validation process with additional layers. These include verification of the business's public phone number, business existence duration, registration number, checks for domain fraud, and authentication of the employment status of the domain owner through telephone calls.

\subsubsection{Webpage Content}
\label{sec:Web_Page_Content}
The other input considered by phishing webpage detection algorithms is webpage content consisting of the HTML code, CSS rules, Javascript codes on tags like \texttt{<a href="$\#$">, <link>, <script>}, and PHP form redirect using \texttt{form action="attackerFile.php"}. When a user enters their credentials into phishing websites, the credentials get redirected to the attacker who then employs these stolen credentials to initiate an identity theft attack, gaining access to legitimate websites on behalf of the user and engaging in malicious activities. Hence the analysis of the webpage content is also an important step for knowing the legitimacy of a webpage.

\subsection{Dataset Repositories}
\label{sec:Dataset_Description}
Phishing dataset repositories play a vital role in training ML classifiers to identify malicious websites and protect users from potential threats. These repositories offer various inputs, such as URLs and webpages, which are essential for training and testing ML models.

These valuable datasets can be found in open-source repositories, and accessible through websites or public sources. However, it is important to note that not all repositories may contain a diverse set of phishing features in their webpages, which can impact the classifier's ability to effectively distinguish between phishing and legitimate webpages.

To ensure the ML classifier can accurately classify new phishing webpages, it is essential to provide it with a diverse dataset. This diversity enables the classifier to learn informative features, enhancing its ability to identify phishing attempts more accurately.

Thankfully, numerous dataset repositories and websites offer both legitimate and phishing samples. These repositories serve as pivotal resources for researchers and developers in their efforts to combat phishing threats. Table~\ref{tab:Dataset_Repositories}, presents an overview of some repositories that contain either legitimate, phishing, or both types of samples, making it easier for practitioners to access and utilize the datasets effectively.


\begin{table}[th]
\centering
\caption{Dataset Repositories}
\label{tab:Dataset_Repositories}
\begin{tabular}{p{2.35cm} C{1.0cm} C{0.9cm} C{0.5cm} C{0.7cm} C{1.0cm}}
\toprule
\multirow{2}{*}{\textbf{Repositories}} & \multicolumn{2}{c}{\textbf{Class}} & \multicolumn{3}{c}{\textbf{Sample Input Type}} \\
\cmidrule(l){2-3}\cmidrule(l){4-6}
& \textit{Legitimate} & \textit{Phishing} & \textit{URLs} & \textit{Content} & \textit{Screenshot} \\ \midrule
    PhishTank~\cite{phishtank} & \ding{51} & \ding{51} & \ding{51} && \ding{51} \\
    UCI~\cite{UCI1} & \ding{51} & \ding{51} & \ding{51} & \ding{51} & \\
    Mendeley~\cite{mendeley} & \ding{51} & \ding{51} & \ding{51} & \ding{51} & \\ 
    Alexa~\cite{alexa_db} & \ding{51} && \ding{51} && \\
    Common Crawl~\cite{common_crawl} & \ding{51} && \ding{51} & \ding{51} & \\
    Tranco~\cite{tranco_db} & \ding{51} && \ding{51} && \\
    DMOZ~\cite{dmoz} & \ding{51} && \ding{51} && \\
    Curlie~\cite{curlie} & \ding{51} && \ding{51} && \\
    OpenPhish~\cite{openphish} && \ding{51} & \ding{51} && \\
    Millersmiles~\cite{millersmiles} && \ding{51} & \ding{51} && \\
    \bottomrule
\end{tabular}
\end{table}

\subsubsection{Legitimate Repositories}
\label{sec:Legitimate_Datasets}
The preprocessing of models commences with the collection of dataset samples from diverse repositories. Several open-source repositories offer legitimate samples comprising URLs, webpage source code, and webpage screenshots. Among these repositories, the following are the most favoured choices for gathering legitimate URLs with webpage source code.
\begin{itemize}
    \item Alexa~\cite{alexa_db} was a dataset repository for legitimate websites that retired on May $1$, $2022$. It contained the top one million domain names and was used as a dataset for phishing webpage detection.
    \item Tanco~\cite{tranco_db} offers a versatile ranking system, designed to cater to various research needs. It compiles data from all available rankings spanning a $30$-day period, ensuring suitability for most research purposes. Additionally, researchers have the flexibility to customize rankings by applying multiple filters to align with their specific requirements. Tranco archives all generated rankings, providing a permanent link to a page that details the methodology behind the ranking and offers the option to download the exact list used in the study. Furthermore, Tranco provides an updated list of the top one million domain names as an alternative option.
    \item DMOZ~\cite{dmoz} was an open-content multilingual directory that was used for the ontology listing of various sites. It is replaced by the Curlie directory~\cite{curlie}. The Curlie directory has a collection of URLs.
    \item Common Crawl~\cite{common_crawl} is a nonprofit organization that maintains and provides access to a large-scale web crawling dataset. This dataset, known as the Common Crawl corpus, contains a comprehensive collection of web pages and their associated metadata, including text, HTML, and other information gathered from websites across the internet. The web data is openly and easily accessible for research, analysis, and various applications.
\end{itemize}

\subsubsection{Phishing Repositories}
\label{sec:Phishing_Datasets}
The dataset collection preprocessing task includes gathering phishing inputs. The datasets listed below consist of phishing URLs marked by cybercrime analysts, along with some URLs raised by users as potential phishing URLs. User-raised phishing URLs undergo analysis based on URL standards features to determine whether they should be classified as phishing or considered legitimate URLs.
\begin{itemize}
    \item OpenPhish~\cite{openphish} provides statistics of the total number of URLs processed and marked as phishing, the brands that are highly targeted to get an insight for developers to perform visual similarity-based approaches to detect a new phishing URL matching to the top brand's webpage. The crucial and helpful information about OpenPhish is that it provides a regular update on the total URLs marked as phishing. The database contains information such as the webpage information, hostname, language supported, IP address presence, country code, SSL certificate, and much more for analysis.
    \item Millersmiles~\cite{millersmiles} contains phishing URLs that occur in a malicious email. It contains $2,636,652$ email scams along with the phishing URLs used in the emails. It is one of the biggest archives of spoof email and phishing scams from several years.
\end{itemize}

\subsubsection{Phishing and Legitimate Repositories}
Several open-source repositories provide inputs for both phishing and legitimate data. These repositories offer essential elements such as URLs, webpage content, and webpage screenshots, which are valuable for the development of efficient models. The following repositories are commonly favoured for their comprehensive inputs covering both classes of data.
\begin{itemize}
    \item PhishTank~\cite{phishtank} is the most commonly used source for collecting phishing URLs and webpage screenshots. It is a community source containing URLs that are verified, validated, marked, and submitted by experts who find a new URL is not safe to surf. It contains a database that can help validate if a new URL falls as phishing or legitimate based on the comparison test with the database URLs. It also provides an API that can help validate if a URL is safe to click and access. Moreover, it helps developers to integrate their work with the PhishTank data to run their applications free of cost.
    \item UCI~\cite{UCI1} is a benchmark database that provides a large collection of features extracted for phishing and legitimate URLs. It contains a total of $30$ features containing URL, content and third-party-based extracted for $11,055$ phishing and legitimate URLs. The dataset is considered for the ML community to enhance their work by considering the given dataset to provide an empirical analytical report.
    \item Mendeley Phishing Websites Dataset~\cite{mendeley}, is a repository containing legitimate and phishing URLs, along with source code, totalling $80,000$ instances. Out of these, $50,000$ instances of legitimate URLs are collected from two sources (namely, Google searches using simple keywords that displayed the top $5$ websites and Ebbu $2017$ Phishing Dataset~\cite{ebbu_2017}) and $30,000$ phishing URLs are collected from three sources (namely, PhishTank~\cite{phishtank}, OpenPhish~\cite{openphish} and PhishRepo~\cite{phishrepo}).
\end{itemize}

\subsection{Features}
\label{sec:Features}
In this section, we discuss a range of valuable characteristics for classifying a given suspicious URL as either phishing or legitimate. This classification relies on the alignment of these characteristics with the feature set outlined in Table~\ref{tab:phishing_detection_feature_classification}. The features used in previous research are organized into four main categories based on URL, webpage content, visual and third-party characteristics.

\begin{table*}[tbh]
\caption{Phishing Webpage Detection Features Classification}
\label{tab:phishing_detection_feature_classification}
\begin{tabular}{cll}
\toprule
\makecell{\textbf{Phishing Features} \\ \textbf{Category}} & \textbf{Features} & \textbf{Description} \\  \midrule \midrule

\multirow{13}{*}{\rotatebox[origin=c]{}{\textbf{URL-based}}} & IP Address & Whether the domain name is replaced by an IP address in the URL \\ 
& URL length & Total character length of URL \\ 
& Subdomains & Number of subdomains levels \\ 
& Presence of \texttt{@} in URL & Browsers ignore anything before \texttt{@} in URL \\ 
& Redirection using \texttt{//} & From Existing Link to intended Destination Link \\ 
& Prefix and Suffix using \texttt{-} & Using the prefix (say, \texttt{login-}, \texttt{secure-}) and suffix (say, \texttt{-logon}, \texttt{-websrc}) \\ 
& Count of Special Characters (\texttt{-}, \texttt{\_}, \texttt{.}) & Count of characters in URL \\ 
& HTTPS in domain name & Domain name contains HTTPS (for example, \texttt{http://https://brand.com}) \\ 
& Suspicious Words & Gaining the trust of users by adding words like: \\&& \quad \texttt{login}, \texttt{websrc}, \texttt{account}, \texttt{secure}, \texttt{siginin} \\
& Obfuscated URL & The IP address in the domain name of URL is replaced in hex or octet format \\
& Self-Signed Certificate & Issuer and Subject fields are same or signed from compromised service\\ 
& Issuer and Domain Owner Information & Contains Certificate Authority and domain owner information \\ \midrule

\multirow{18}{*}{\rotatebox[origin=c]{}{\makecell{\textbf{Webpage} \\ \textbf{Content-based}}}} & Favicon link & \texttt{<link>} tag containing favicon link with \texttt{.ico} extension \\ 
& Number of Links Pointing to Page & Count of total number of \texttt{<a href="">} containing the domain in URL \\ 
& URL of Anchor & Anchor Tags (\texttt{<a>}) containing following patters: \\&& \quad \texttt{<a href="">}, or \texttt{<a href="\#">}, or \texttt{<a href="\#content">}, or \\ && \quad \texttt{<a href="Javascript::void(0);">} \\ 
& Domain Name Frequency in Anchor tag & Count of total Anchor tags containing the domain name in the \texttt{href} URL \\ 
& Links in Script Tags & Links like \texttt{<meta>}, \texttt{<link>}, \texttt{<script>} tags used for manipulating content \\ 
& \texttt{form action=""} field value & The \texttt{action} field in the \texttt{form} tag contains the redirection of input information to \\&& \quad attackers' mail or gets stored in the attacker's owned database\\ 
& Hidden Webpage Content & Phishing webpages hide a few \texttt{<div>}, \texttt{<input>}, \texttt{<button>} tags using: \\&& \quad \texttt{<div style:"visibility:hidden">} or \\&& \quad \texttt{<div style:"display:none"} or \\&& \quad \texttt{<button disabled="disabled"} or \\&& \quad \texttt{<input disabled="disabled">} \\
& IFrame & Presence of total \texttt{<iframe>} tags in webpage \\
& Using onMouseOver to hide links & Hide the links through Javascript command \\ 
& Disabling Right Click & Not allowing targets to view the source code \\
& Using pop-up window & Pop-up window for content input \\ \midrule


\multirow{3}{*}{\rotatebox[origin=c]{}{\textbf{Visual-based}}} & Webpage Screenshot & Analysis of the webpage screenshot to locate the layout of different fields \\ 
& Domain Logo & Analysis of Logo for similarity with legitimate logo \\
& Favicon Image & Analysis of suspicious webpage favicon with legitimate favicons \\ \midrule

\multirow{6}{*}{\rotatebox[origin=c]{}{\textbf{Third-party-based}}} & Domain Registration Length & Owners make payment for the domain registration of a website \\ 
& Age of Domain & If less than $6$ months then maybe phishing site \\ 
& DNS Record & Provides corresponding IP address domain name \\ 
& Website Traffic & Visited users list on a website \\ 
& Page Rank & Rank of the website on Google search engine \\ 
& Google Index & Adding webpages on Google search \\ \bottomrule
\end{tabular}
\end{table*}

\subsubsection{URL}
\label{sec:URL_as_Input_with_Features}
The URL-based feature examines various URL fields, such as the domain name, the protocol, subdomains, path and much more considering just the URL as a string. Through observational studies, researchers have examined the characteristics features that differentiate between a phishing URL with a legitimate URL. Table ~\ref{tab:phishing_detection_feature_classification} presents a set of URL-based features such as the presence of an IP address in place of a domain name, the existence of the \texttt{@} symbol, the presence of HTTPS in the domain name, count of special characters, URL length, the number of subdomains, and several other features. These features are utilized to differentiate between legitimate and phishing URLs by assigning binary values ($0$ or $1$) based on the presence or absence of a particular feature or by assessing a count against predefined threshold values. Consequently, this category of features is particularly beneficial for approaches that require a limited number of features and aim to avoid visiting the actual phishing webpage.

\subsubsection{Webpage Content}
\label{sec:Webpage_Content_as_Input_with_Features}
Webpage content-based features serve a vital role in conducting a more comprehensive analysis of suspicious webpages, going beyond the examination of URL-based characteristics. In Table~\ref{tab:phishing_detection_feature_classification}, you can find a list of the most frequently observed features that aid in distinguishing phishing webpages from legitimate ones. Attackers typically manipulate the outlinks, which are resources within webpages, by tampering with HTML tags like \texttt{<a>}, \texttt{<link>}, and \texttt{<script>}, diverting these links to different webpages. Additionally, attackers utilize CSS properties like \texttt{visibility:hidden} and \texttt{display:none} to conceal specific elements on a legitimate webpage and replace them with malicious content.

The primary objective of these attackers is to entice users and harvest their sensitive information. To achieve this, they often modify the \texttt{<form action="">} field value, redirecting user data either to the attacker's email via \texttt{mailto:} or storing it in a database owned by the attacker using \texttt{filename.php}. Several other features, such as disabling right-click functionality and utilizing \texttt{onmouseover}, are employed to prevent users from easily inspecting and viewing the source code of the suspicious webpage.

\subsubsection{Visual}
\label{sec:Visual_as_Input_with_Features}
Modern phishing webpage detection methods now take into account the visual presentation of a webpage, examining elements such as the header, forms, images, logos, and other design aspects to enhance the accuracy of webpage classification as either phishing or legitimate. Table~\ref{tab:phishing_detection_feature_classification} outlines three primary visual-based features that are frequently employed: the webpage screenshot, favicon image, and the domain logo. These images are subjected to a comparison process with established legitimate webpages to calculate a similarity score, which is then used to determine whether the examined webpage is suspicious or legitimate.

\subsubsection{Third-party}
\label{sec:Third_Party_as_Input_with_Features}
When a webpage is hosted, it typically provides various details such as the registration date, registrar name, domain age, PageRank, Google Index, and other relevant information. These attributes are categorized as third-party-based features and are typically acquired through the use of the WHOIS~\cite{whois} service, or using the Registration Data Access Protocol~\cite{RDAP}. In Table~\ref{tab:phishing_detection_feature_classification}, you can find a description of commonly employed third-party-based features that play a significant role in the classification of a suspicious URL as either phishing or legitimate. For instance, domain age is a specific feature that reveals the length of time the webpage has been hosted.


\subsection{Feature Selection Algorithms}
\label{sec:Feature_Selection_Algorithms}
Feature selection algorithms in ML are methods that pinpoint the most informative and pertinent feature subsets from a larger pool of available features. The main goal of feature selection is to optimize the model's performance, diminish over-fitting, enhance computational efficiency, and promote interpretability by singling out only the most informative features that significantly influence the target variable or prediction task~\cite{liu1998feature, sarker2020context}. Moreover, these algorithms offer a list of features that can be omitted due to their correlation with other features.
\begin{enumerate}
    \item \textit{Information Gain (IG)~\cite{pedregosa2011scikit}} is a feature selection criterion used mainly in decision tree-based algorithms. It measures the reduction in uncertainty (entropy) of the target variable after considering a particular feature. Features with higher IG are preferred for splitting nodes in decision trees, as they contribute more to the classification task by effectively discriminating between different classes.
    
    \item The \textit{Correlation-based Feature Selection (CFS) Algorithm~\cite{michalak2006correlation}} ranks features according to their correlation with the target class. In phishing webpage detection, numerous features are often considered to enhance the model's robustness in classifying suspicious URLs. However, some features in the selected set may have a high correlation, leading to redundant information. CFS algorithms can identify such highly correlated features, allowing the approach to choose one of them and reduce computation time.

    \item \textit{Fuzzy Rule Set (FRS)-based Algorithm~\cite{dubois1990rough}} selects the most relevant features from the feature set. The feature set contains labels marked for each of the features and thus fuzzy logic provides weights on each feature to know the most weighted features that effectively help make the detection more accurate for new inputs.
\end{enumerate}

Thus, feature selection algorithms are essential tools in ML to choose the most relevant and informative features, enhancing model performance, interpretability, and efficiency. IG, CFS, and FRS are specific feature selection algorithms, each with its unique approach to selecting relevant features.

In ML, various feature engineering techniques are employed to assess the informativeness of selected features and their contribution to model training, ultimately leading to improved performance. The choice of feature engineering techniques depends on the data characteristics and the specific ML algorithm being used. Below, we discuss some common feature engineering techniques:
\begin{enumerate}
    \item Term Frequency-Inverse Document Frequency (TF-IDF) Algorithm~\cite{jalilifard2021semantic}: is a numerical representation used to measure the importance of a term (\textit{t}) within a document (\textit{d}) relative to a collection of documents (corpus, \textit{D}). In the context of phishing webpage detection, TF-IDF is commonly employed in text mining and information retrieval tasks to assess the significance of specific words in URLs and webpage content compared to their frequency across the entire dataset.

    The algorithm computes the TF (Term Frequency) and IDF (Inverse Document Frequency) for each term \textit{t}, document \textit{d}, and the corpus \textit{D} using the equations:

       - Term Frequency (TF):
         \begin{equation}
         \label{eq:TF}
         \begin{split}
             \text{TF}(t, d) &= \frac{f_{t, d}}{\sum_{t' \in d}^{} f_{t', d}}
         \end{split}
         \end{equation}
         
       - Inverse Document Frequency (IDF):
         \begin{equation}
         \label{eq:IDF}
             \text{IDF}(t, D) = \log_{e} \left( \frac{|D|}{|{d \in D : t \in d}|} \right)
         \end{equation}
    
    The TF-IDF value is then obtained by multiplying the TF and IDF values:
    
    \begin{equation}
     \label{eq:TF-IDF}
     \text{TF-IDF} = \text{TF}(t, d) \times \text{IDF}(t, D)
    \end{equation}
    

    \item Principal Component Analysis (PCA)~\cite{abdi2010principal} is a dimensionality reduction technique used to transform the original features of a dataset into a new set of orthogonal (uncorrelated) features called principal components. These components capture the most significant sources of variance in the data and can be used to reduce noise and redundancy. In the phishing webpage detection domain, PCA helps retain information from the most correlated features in the dataset.
\end{enumerate}

\subsection{ML Algorithms}
\label{sec:ML_algorithms}
The phishing webpage detection problem revolves around employing a classification-based approach, wherein a training dataset is provided with accurate class labels for each URL, designating them as either phishing or legitimate. This makes it a supervised learning approach, as each training input is associated with a specific class label. Supervised learning encompasses two main categories: classification and regression. In the context of this problem, the focus is on classification, which involves the process of identifying, understanding, and grouping objects into pre-defined classes. ML algorithms used for classification are trained on the provided dataset to gain the ability to predict the likelihood of new input falling into one of the predefined classes, thereby enabling effective phishing webpage detection. In the realm of ML, various classifiers are categorized and defined as follows:
\begin{enumerate}
    \item \textit{Decision Trees:} This group includes RF (Random Forest), DT (Decision Tree), J48 (C4.5 in WEKA), CART (Classification and Regression Trees), and BART (Bayesian Additive Regression Trees).
    \begin{itemize}
        \item RF~\cite{breiman2001random} is an ensemble learning method that builds multiple decision trees during training and outputs the mode of the classes for classification tasks or the mean prediction for regression tasks.
        \item DT~\cite{quinlan2014c4} is a tree-like model that recursively splits the data based on feature values, with each internal node representing a decision and each leaf node representing a class label or a regression value.
        \item J$48$ (C$4.5$ in WEKA)~\cite{quinlan2014c4} is an implementation of the C$4.5$ algorithm used for constructing decision trees from a dataset. It utilizes information gain to determine the best attribute for node splitting.
        \item CART~\cite{gordon1984classification} is a decision tree algorithm that can be used for both classification and regression tasks. It uses the Gini impurity for classification and mean squared error for regression as the criteria for node splitting.
        \item BART~\cite{bharadiya2023review} is a Bayesian ML method that uses a sum of decision trees to model the relationship between input features and output values.
    \end{itemize}

    \item \textit{Neural Networks:} This category comprises LSTM (Recurrent Neural Network), MLP (Multi-Layer Perceptron), DNN (Deep Neural Network), RNN (Recurrent Neural Network), CNN (Convolutional Neural Network), and ELM (Extreme Learning Machine).
    \begin{itemize}
        \item RNN~\cite{goodfellow2016deep} is a type of neural network designed to handle sequential data by maintaining hidden states to capture temporal dependencies.
        \item LSTM~\cite{goodfellow2016deep} is a type of recurrent neural network (RNN) that is capable of learning long-term dependencies in sequential data, making it suitable for tasks like natural language processing and time series prediction.
        \item MLP~\cite{pedregosa2011scikit} is a feedforward neural network with one or more hidden layers between the input and output layers, used for a wide range of tasks, including classification and regression.
        \item DNN~\cite{sharma2017era} refers to any neural network with multiple hidden layers, allowing it to learn hierarchical representations of data.
        \item CNN~\cite{lecun1998gradient} is a type of neural network commonly used for image and video analysis, employing convolutional layers to automatically learn spatial hierarchies of features.
        \item ELM~\cite{huang2012semi} is a feedforward neural network with randomly initialized hidden layers, known for its fast training and good generalization performance.
    \end{itemize}

    \item \textit{Support Vector Machine (SVM):} SVM and SMO (Sequential Minimal Optimization) fall under this classification.
    \begin{itemize}
        \item SVM~\cite{keerthi2001improvements} is a supervised learning algorithm used for classification and regression tasks. It finds an optimal hyperplane that best separates data points of different classes or predicts a continuous value for regression tasks.
        \item SMO~\cite{huang2015sequential} is an algorithm used to efficiently solve the quadratic programming problem associated with training an SVM.
    \end{itemize}

    \item \textit{Ensemble Learning:} Adaboost, PCA-RF (Ensemble Learning with Principal Component Analysis), and XGboost (Gradient Boosting) are part of this ensemble learning approach.
    \begin{itemize}
        \item Adaboost~\cite{pedregosa2011scikit} is an ensemble learning method that iteratively combines weak classifiers to create a strong classifier, giving higher weight to misclassified samples in each iteration.
        \item PCA-RF~\cite{gupta2022pca} combines random forests with principal component analysis (PCA) to improve the performance of random forests, particularly on high-dimensional data.
        \item XGboost~\cite{pedregosa2011scikit} is a popular gradient-boosting framework that builds multiple weak learners in a sequential manner, each correcting the errors of its predecessors.
    \end{itemize}

    \item \textit{Rule-Based Classifiers:} This category encompasses RIDOR (RIght direction OR), OneRule, and Conjunctive rule.
    \begin{itemize}
        \item RIDOR~\cite{witten2002data} is a rule-based classifier that employs the OR operation to combine multiple rule conditions.
        \item OneRule~\cite{holte1993very} is a simple rule-based classifier that selects a single rule based on the feature that maximizes accuracy.
        \item Conjunctive rule~\cite{pouriyeh2017comprehensive} is a type of rule-based classifier that forms rules using conjunctions of feature conditions.
    \end{itemize}

    \item \textit{Linear Classifiers:} LR (Logistic Regression)is an example of a linear classifier.
    \begin{itemize}
        \item LR~\cite{cessie1992ridge} is a linear classifier used for binary classification tasks. It predicts the probability of an instance belonging to a certain class using a logistic function.
    \end{itemize}

    \item \textit{Probabilistic Classifier:} NB (Naive Bayes) is a representative of this type.
    \begin{itemize}
        \item NB~\cite{john2013estimating} is a probabilistic classifier based on Bayes' theorem, assuming independence between features to calculate the likelihood of a class given the features.
    \end{itemize}

    \item \textit{Instance-Based (Lazy) Learning:} The well-known $k$NN ($k$-Nearest Neighbors) method belongs to this group.
    \begin{itemize}
        \item $k$NN~\cite{aha1991instance} is a lazy learning algorithm that classifies instances based on the majority class of their $k$-nearest neighbours in the feature space.
    \end{itemize}

    \item \textit{Evolutionary Algorithm:} e-DRI (evolutionary Decision Rule Induction) is an example of a classifier based on evolutionary algorithms.
    \begin{itemize}
        \item e-DRI~\cite{freitas2002data} is a classifier based on evolutionary algorithms, designed to discover and optimize decision rules from data.
    \end{itemize}
\end{enumerate}

These classifiers are defined and utilized in various ML tasks to facilitate pattern recognition and decision-making processes.


\subsection{DL Algorithms}
\label{sec:DL_Algorithms}
ML classifiers necessitate the process of feature engineering, wherein relevant features must be manually crafted from the training dataset to enable effective phishing webpage detection. In contrast, DL algorithms do not demand such feature engineering, as the algorithms autonomously extract relevant features from the provided training dataset. Moreover, DL algorithms can autonomously ascertain the hierarchical features within an unsupervised dataset, encompassing webpage content and images. In the existing DL-based phishing webpage detection approaches the used algorithms are recurrent neural networks (RNNs), convolutional neural networks (CNNs), long short-term memory networks (LSTMs), and deep belief networks (DBNs). These DL algorithms are described as follows:
\begin{enumerate}
    \item Recurrent Neural Networks (RNNs)~\cite{halgavs2020catching} are neural networks designed for processing sequential data like text, speech, and time series. The connections within the neurons in the network allow the algorithm to maintain a hidden state to capture previous input information. RNNs are thus considered suitable for processing the URLs, and webpage content in phishing detection approaches.
    \item Convolutional Neural Networks (CNNs)~\cite{o2015introduction} consists of two layers namely, convolutional and pooling layers. A CNN architecture is formed by stacking these layers. CNNs are used for image processing and grid-based data and are also applied in time series and sequential data. CNNs are used in various domains such as image classification, object detection and image segmentation. One of these applications is the analysis of the webpage screenshots and logos in phishing webpage detection approaches.
    \item Long Short Term Memory (LSTM)~\cite{van2020review} are RNNs designed to overcome the vanishing gradient problem. LSTMs have a more complex structure in comparison to the RNNs, consisting of specialized gates to control the flow of information between the networks. LSTMs are mainly used in speech recognition, generating the next characters based on the previous processed characters where long-range dependencies are necessary. Thus, LSTMs are used in URL-based phishing detection approaches.
    \item Deep Belief Networks (DBNs)~\cite{hinton2006fast} is developed on two types of neural networks namely, Belief Networks and Restricted Boltzmann Machines (RBMs)~\cite{hua2015deep} to manage large data by including hidden layers. DBNs are pre-trained layer-by-layer as RBMs before fine-tuning the entire network. DBNs are helpful in phishing webpage detection approaches due to their capacity for effectively learning hierarchical data representations, extracting pertinent features, and acquiring insights into the underlying structures of both legitimate and phishing webpages through unsupervised pre-training on extensive datasets, among other advantages.
\end{enumerate}

\subsection{Performance Metrics}
\label{sec:Performance_Metrics}
In the context of performance metrics in ML, classifiers are trained on input datasets containing samples from multiple classes. Features are then extracted and selected to prepare the model for real-world sample detection. The model's performance is a measure of its ability to accurately classify new samples into their respective classes.

In ML, classifier performance metrics are computed using the four statistical values from the confusion matrix~\ref{tab:Confusion_Matrix}: True Positive (TP), True Negative (TN), False Positive (FP), and False Negative (FN). These values play a fundamental role in determining the accuracy, precision, recall, and F1 score of the model.

\begin{table}[htb]
\centering
\caption{Confusion Matrix}
\label{tab:Confusion_Matrix}
\begin{tabular}{@{}cccc@{}}
\toprule
\multicolumn{2}{l}{\multirow{2}{*}{}} & \multicolumn{2}{c}{\textbf{Predicted}} \\ \cmidrule(l){3-4} 
\multicolumn{2}{l}{} & \textit{Positive (1)} & \textit{Negative (0)} \\ \cmidrule(r){1-2}
\multirow{2}{*}{\makecell{\textbf{Actual}}} & \textit{Positive (1)} & True Positive & False Negative \\
 & \textit{Negative (0)} & False Positive & True Negative \\ \bottomrule
\end{tabular}
\end{table}

The terms in the confusion matrix are defined as follows:
\begin{description}
    \item True Positive (TP): This occurs when the model correctly predicts a data point as positive (e.g., having a disease) and the actual class of the data point is also positive.
    
    \item True Negative (TN): This happens when the model correctly predicts a data point as negative (e.g., not having a disease) and the actual class of the data point is also negative.
    
    \item False Positive (FP): This occurs when the model incorrectly predicts a data point as positive, but the actual class of the data point is negative.
    
    \item False Negative (FN): This happens when the model incorrectly predicts a data point as negative, but the actual class of the data point is positive.
\end{description}
    
For instance, in a medical diagnosis scenario: TP would be when the model correctly predicts that a patient has a disease (and the patient actually has the disease).
TN would be when the model correctly predicts that a patient does not have a disease (and the patient does not have the disease).
FP would be when the model predicts a patient has a disease (but the patient does not have the disease).
FN would be when the model predicts a patient does not have a disease (but the patient actually has the disease).

These definitions of TP, TN, FP, and FN are necessary for evaluating the performance metrics like accuracy, precision, recall (sensitivity), specificity, and F1 score of ML models in various classification tasks.

The confusion matrix produces a set of terms based on which the following metrics are defined.
\begin{description}
    \item Accuracy measures the percentage of correct predictions made by a model out of all possible predictions in classification problems. It is computed as:
    
    \[ Accuracy = \frac{\text{TP} + \text{TN}}{\text{TP} + \text{FP} + \text{TN} + \text{FN}} \]
    \item Precision is the proportion of true positives (correctly predicted positives) to the total number of predicted positives. It is calculated as:
    
    \[ Precision = \frac{\text{TP}}{\text{TP} + \text{FP}} \]
    \item Recall (Sensitivity) is the proportion of true positives to the total number of actual positives. It is given by:
    
    \[ Recall = \frac{\text{TP}}{\text{TP} + \text{FN}} \]
    \item Specificity measures true negatives over the total number of actual negatives. It is the opposite of Recall and is computed as:
    
    \[ Specificity = \frac{\text{TN}}{\text{TN} + \text{FP}} \]
    \item F1 Score is the harmonic mean of precision and recall, capturing a balance between the two metrics. It is calculated as:
    
    \[ \text{F1 Score} = \frac{2 \times Precision \times Recall}{Precision + Recall} \]
\end{description}

Accuracy is suitable when the data balance between classes is almost equal, but for imbalanced datasets, precision, recall, and F1 score are more informative metrics to evaluate model performance. These metrics provide valuable insights into the model's effectiveness in classification tasks and guide data scientists in making informed decisions about the model's performance and potential improvements.

\section{Phishing Webpage Detection Methodologies}
\label{sec:Methodology_Taxonomy_for_Phishing_Detection}
Phishing webpage detection is possible in two methods: training the users in identifying and notifying the URLs as phishing and automated web phishing webpage detection. In the method involving users to identify the sites as legitimate or phishing, they have to be knowledgeable about the parameters that are important to observe in the webpage to mark it as either a legitimate page or a phishing page. However, this method of user training is not sufficient since the attackers can develop new features and alternate approaches to lure the users. Considering this, automated phishing webpage detection techniques are developed to protect users from falling for phishing attacks. In the automated phishing webpage detection method, models are built based on the features that can be identified on a phishing webpage. The features become a governing factor in classifying the webpage as phishing or legitimate.

The phishing webpage detection approaches can be categorized as \textit{URL-based evaluation}, \textit{webpage content-based evaluation}, and \textit{hybrid approach} based on the inputs considered for classifying the suspicious URL as phishing or legitimate. The phishing webpage detection approaches are categorized as shown in Figure~\ref{fig:Taxonomy_on_Phishing_Detection_Methodology}:

\begin{figure}[th]
\begin{center}
\begin{forest}
    for tree = {
        draw,
        grow=0,
        edge={->,>=latex},
        s sep=2mm,
        inner sep=4,
        anchor=parent,
        forked edges,
    }
    [\rotatebox{90}{\textbf{Phishing Webpage Detection Methodologies}}
        [Hybrid Approaches
           [\cite{rao2019detection, somesha2020efficient}
            ]
        ]
        [\rotatebox{90}{Webpage-based Approaches}
            [DL-based Techniques on \\ Webpage Screenshots
                [\cite{yi2018web, abdelnabi2020visualphishnet}]
            ]
            [Webpage Screenshot-based \\ Techniques
                [\cite{medvet2008visual, fu2006detecting, afroz2011phishzoo, lin2021phishpedia}
                ]
            ]
            [ML-based Techniques
                [\cite{jain2019machine, pan2006anomaly, garera2007framework, zhang2017two}
                ]
            ]
            [Content Similarity \\ Techniques
                [\cite{mao2017phishing, zhang2007cantina, xiang2011cantina+, hara2009visual}
                ]
            ]
        ]
        [\rotatebox{90}{URL-based Approaches}
            [DL-based Techniques
                [\cite{le2018urlnet}
                ]
            ]
            [ML-based Techniques
                [\cite{kumar2020phishing, sharma2019phishalert, jain2018phish, alam2020phishing, abdelhamid2017phishing, sonmez2018phishing, verma2015character, rao2020catchphish, shirazi2018kn0w, sahingoz2019machine, jeeva2016intelligent}
                ]
            ]
            [Heuristic Techniques
                [\cite{ma2009beyond, blum2010lexical}
                ]
            ]
            [Certificate-based Techniques
                [\cite{torroledo2018hunting}
                ]
            ]
            [List-based Techniques
                [\cite{cao2008anti, wang2008light, teraguchi2004client}
                ]
            ]
        ]
    ]
\end{forest}
\caption{Phishing Webpage Detection Methodologies}
\label{fig:Taxonomy_on_Phishing_Detection_Methodology}
\end{center}
\end{figure}
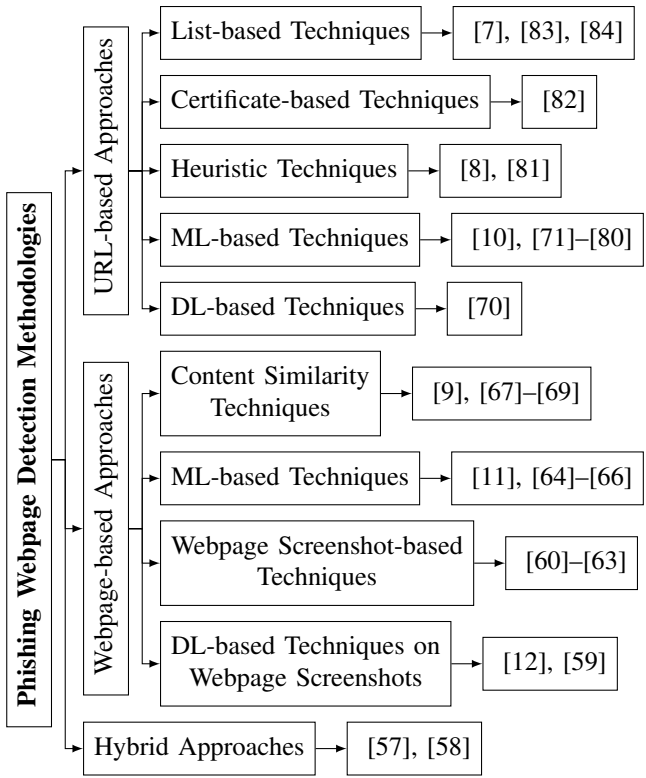


\subsection{URL-based Approaches}
\label{sec:URL-based_Approach}
The phishing webpage detection is based on the input parameter as the URL contains the features restricting to the URLs including a few as the presence of IP address, presence of \texttt{@} symbol, count of \texttt{-} in the URL path, and much more. This category also includes certificate-based features such as issuers' and domain owner's details in the webpage certificate. In addition to this, the comparison is also made through a string comparison of the suspicious input URL with the list of URLs present in the whitelist and blacklist dataset. These categories of the URL-based evaluation can be clubbed along with the ML and DL-based approaches for better accuracy values and also to decrease the computation time.

\begin{figure*}[th]
    \centering
    \includegraphics[width=1.0\textwidth]{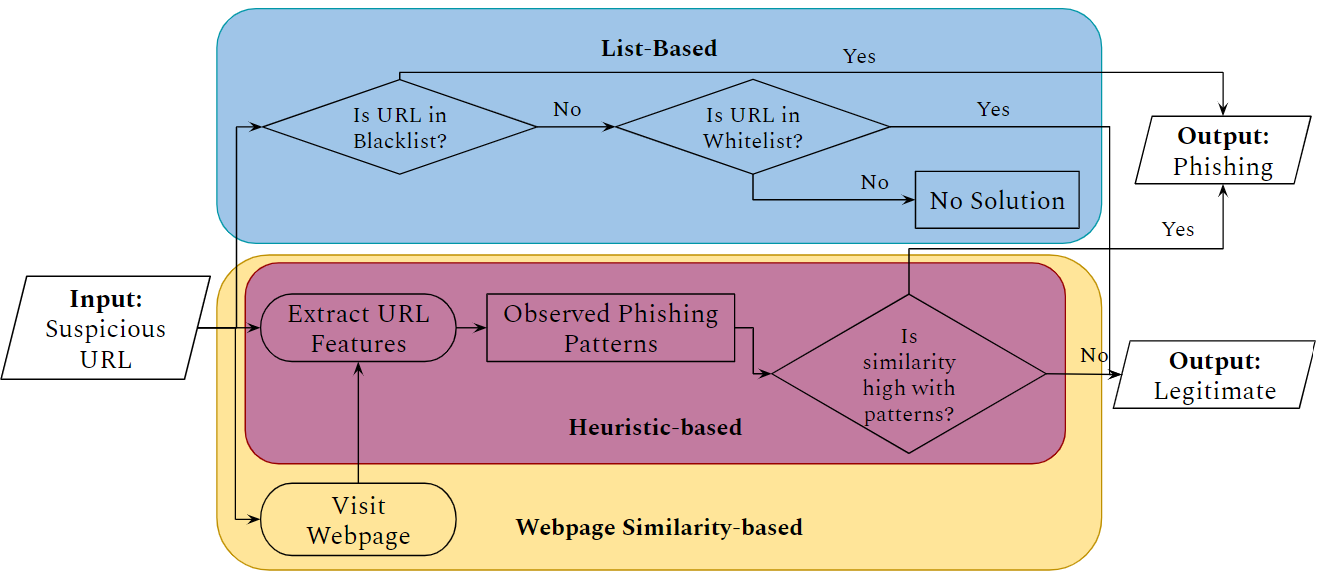}
    \caption{List, Heuristic and Webpage Content Similarity-based Phishing Detection Approaches}
    \label{fig:list_heuristic_webpage_content_based}
\end{figure*}

\subsubsection{List-based Techniques}
This technique involves preventing systems from accessing fraudulent URLs by a two-pronged approach called \emph{whitelisting} and \emph{blacklisting}. The two-pronged approach helps in preventing a system from accessing URLs that land on a malicious webpage, and also email attachments that might contain malware. The whitelist contains a list of legitimate URLs, and the blacklist contains a list of phishing URLs. A suspicious input URL is matched with one of the classes (phishing or legitimate) by comparing with the whitelist and blacklist through a string-matching-based algorithm. Figure~\ref{fig:list_heuristic_webpage_content_based} describes the flow of the list-based approach on an input URL and the series of steps it goes through to get marked as legitimate (gets added to the whitelist database) or phishing (as a warning message). However, list-based phishing detection techniques do not detect zero-day phishing.


Wang \textit{et al.}~\cite{wang2008light} and Chou \textit{et al.}~\cite{teraguchi2004client} utilize the whitelist-based approach for URL classification. The browsers such as Google browser, Mozilla Firefox, and Safari use the Google Safe Browsing (GSB)~\cite{GSB} service for URL classification based on the whitelist-based approach. However, the GSB service cannot detect the new phishing URLs which are called zero-day phishing sites since these sites are not yet marked as phishing and have not yet been added to the blacklist. There are online tools such as Netcraft~\cite{netcraft}, Avast~\cite{avast}, Quick Heal~\cite{quickheal}, and McAfee~\cite{mcafee} that help in the classification of phishing URLs on list-based approaches. But, Netcraft does not raise an alert for a phishing URL if it is not found to be $100\%$ phishing. Hence, the researchers proposed various other approaches for detecting zero-day phishing sites.

\subsubsection{Certificate-based Techniques}
The certificate-based approach helps in identifying various features from the certificate of a given webpage. A webpage when hosted contains a padlock icon in the address bar of a browser with an HTTPS protocol which represents that the webpage uses an encryption standard on the input information provided on the webpage in form fields. Over the years, users have understood that the webpage containing the HTTPS protocol in its URL is encrypted and trustworthy to get connected from their systems. However, attackers have developed self-signature standards in order to populate their webpages that contain HTTPS and a padlock icon.

Torroledo \textit{et al.}~\cite{torroledo2018hunting} proposed a DNN for determining malicious certificates using features that are extracted from the content of the certificate. The authors performed a survey to extract public reviews on the presence of the secure symbol and HTTPS flag and they found from the survey that only three-quarters of the responses believed that the symbol is meant for encryption which is the true reason. The experiment analysis found that phishing sites usually use self-signed certificates to steal the users’ information. The DNN was trained on the dataset from Vaderetro~\cite{vaderetro}, abuse.ch~\cite{abuse.ch} and censys.io~\cite{censys.io} for malicious certificates and the legitimate certificates from Alexa~\cite{alexa_db}. The experimental results show that the DNN model shows an accuracy of $94.87\%$ for malicious certificates and $88.64\%$ for phishing certificates.

\subsubsection{Heuristic Techniques}
Due to the limitations of the traditional list-based technique in detecting zero-day phishing attacks, researchers have introduced a novel approach for categorizing suspicious URLs as either phishing or legitimate. This approach relies on scrutinizing the URL fields and extracting observations that reveal specific patterns distinguishing phishing URLs from legitimate ones. The process of classifying a suspicious input URL as phishing or legitimate is depicted in Figure~\ref{fig:list_heuristic_webpage_content_based}. Within this technique, a given suspicious URL undergoes an analysis of its URL features, which aids in determining its classification. If the majority of patterns found in the suspicious URL align with the observed patterns, it is identified as either phishing or legitimate. As a result, this approach effectively tackles the shortcomings of the list-based technique.

Ma \textit{et al.}~\cite{ma2009beyond} introduced a method for distinguishing between phishing and legitimate URLs by leveraging lexical and host features, yielding an impressive classification accuracy of approximately $95\%$. The lexical features involve employing bigrams, which serve as valuable indicators to identify distinct characteristics in URLs. On a similar note, Blum \textit{et al.}~\cite{blum2010lexical} put forth an online learning technique for URL classification, focusing solely on lexical features while neglecting host-based features. Remarkably, this online learning strategy achieves an even higher accuracy rate of $97\%$.


\subsubsection{ML-based Techniques}
\begin{figure*}[h]
    \centering
    \includegraphics[width=1.0\textwidth,]{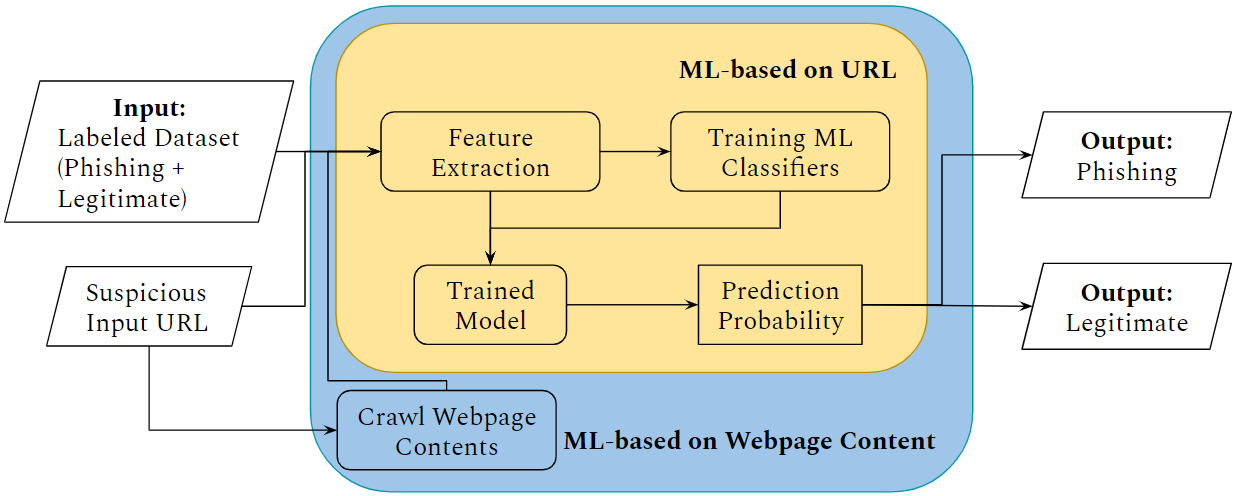}
    \caption{ML Approach on URL and Webpage}
    \label{fig:ml_url_content_based}
\end{figure*}

The URL-based phishing detection approach can be extended with the application of ML-based techniques which are quite useful in extracting the features from the input training dataset and help classify the new suspicious URLs as either phishing or legitimate. Thus, this category in URL-based evaluation is a technical implementation using various ML algorithms. Figure~\ref{fig:ml_url_content_based} shows the technical approach for feature extraction and classification through ML classifiers. The ML classifiers are trained on the dataset by extracting relevant features to classify zero-day URLs as phishing or legitimate. RF, DT, SVM, and NB are the most commonly used ML classification algorithms, resulting in higher prediction. There are numerous research proposals and implementations in this category because the features are extracted only from the URL and certificate and the webpages are not visited for the computation. This optimizes the time of computation and also helps in faster classification.


Rakesh \textit{et al.}~\cite{verma2015character} proposed a robust classification technique for proving the legitimacy of the input URL by considering only a few features that are run on four different datasets namely: Alexa~\cite{alexa_db} and DMOZ~\cite{dmoz} for legitimate URLs and PhishTank, Huawei Digital and APWG for phishing URLs provided as data to the ML classifiers: RF, PART, SMO, LR, NB and J$48$ algorithms. The features considered in the study include Kolmogorov-Smirnov distance, Kullback-Leibler Divergence, Euclidean Distance, character frequencies, edit distance, length of URL, \texttt{@} and \texttt{-} symbols, count of punctuation symbols, number of TLDs in URL, number of target words, IP address and number of suspicious words. The authors also performed diversified training and testing on different datasets in five folds to verify the performance of the classifiers. The experimental results show that the RF algorithm obtains an accuracy of $96.22\%$ with an FP rate of $3.61\%$ which achieves a better performance as compared to other ML algorithms considered in the study for phishing URL classification.

Abdelhamid \textit{et al.}~\cite{abdelhamid2017phishing} introduced a technique called Enhanced Dynamic Rule Induction (eDRI) for detecting phishing attacks. The study included 11000 instances collected from PhishTank~\cite{phishtank} and MillerSmiles~\cite{millersmiles} which are marked either legitimate or phishing. The authors claimed that eDRI is the first rule induction-based phishing webpage detection model. The technique considered the Remove Replace Feature Selection Technique (RRFST) to reduce the dimensionality of the input dataset. A comparative analysis is done by considering ML algorithms like C$4.5$, SVM, AdaBoost, eDRI, OneRule, Conjunctive rule, and RIDOR and eDRI outperformed all the other models except C$4.5$ in achieving better accuracy values. C$4.5$ uses the IG algorithm resulting in its higher accuracy than eDRI at the cost of generating $272$ more rules. Thus eDRI is as accurate as C$4.5$ with fewer rule generation in classifying URLs.

Sonmez \textit{et al.}~\cite{sonmez2018phishing} proposed a classification model for phishing webpage detection considering $30$ features selected from the UC Irvine repository~\cite{UCI1}. The approach considered NN for faster computations to generalize the features and outperformed SVM and NB classifiers. The paper describes the heuristics of the $30$ features helpful in classifying the input URL as phishing or legitimate. The ELM model achieves an accuracy of $95.34\%$ outperforming NB (with accuracy $93.80\%$) and SVM (with accuracy $92.98\%$) classifiers.

Ankit \textit{et al.}~\cite{jain2018phish} proposed a URL-based feature for phishing webpage detection. For the study, the authors have considered $14$ URL-based features namely: IP address in URL, subdomain, presence of \texttt{@} symbol, number of hyphens \texttt{-} in the URL, length of URL, suspicious words in URL, position of top-level-domain, the occurrence of \texttt{//} in URL, HTTPS protocol, number of times HTTPS occurs, DNS lookup, inconsistent URL, age of the domain.  The dataset contains $32,951$ phishing URLs from PhishTank~\cite{phishtank} and $2,500$ legitimate URLs from DMOZ~\cite{dmoz} trained on two ML algorithms NB and SVM. The experiment analysis is performed in two portions of datasets, the first portion containing $15,000$ total URLs ($14,000$ phishing and $1,000$ legitimate) that achieved an accuracy value of $64.74\%$ for the NB classifier and $76.04\%$ for the SVM classifier. The second portion contains $25,000$ URLs ($23,000$ phishing and $2,000$ legitimate) with an accuracy of $76.87\%$ for NB and $91.28\%$ for SVM. The limitation of this work is the biased selection of the dataset which means that the total number of URLs selected for training and testing is not in the same proportion which leads to improper feature vector generation skewing to the larger class.

Shirazi \textit{et al.}~\cite{shirazi2018kn0w} introduced the first approach for phishing classification based on domain-name-based features. The work aimed to perform the classification using a biased dataset for which $1,000$ URLs are selected from PhishTank, $2013$ phishing URLs from OpenPhish, and $1,000$ legitimate URLs from Alexa with a ratio of $80:20$ for training and testing. The work involved classification by considering only eight features (HTTPS present, domain length, page title match, frequency of domain name, non-alphabetical characters, copyright logo match, link ratio in the body, link ratio in body, and URL length) that extracted a feature vector for six ML classifiers namely, SVM linear, SVM Gaussian, Gaussian NB, $k$NN, DT, Gradient Boosting. The experiment analysis was carried out in two folds of the dataset. The first fold included a dataset from Alexa~\cite{alexa_db} $+$ PhishTank~\cite{phishtank} and the second fold included a dataset from Alexa~\cite{alexa_db} $+$ PhishTank~\cite{phishtank} $+$ OpenPhish~\cite{openphish}. The result proved that URL length is a major feature to consider for obtaining the best accuracy values and in this work, $k$NN and Gradient Boosting algorithms achieved an accuracy of $99.7\%$ and $99.75\%$ respectively.

Kumar J. \textit{et al.}~\cite{kumar2020phishing} proposed an ML approach for phishing URL detection in this paper. The URL preprocessing begins with selecting the URL features that are generally classified as URL lexical structure-based features, Domain name-related features, and Page-based features. The feature extraction phase of the algorithm considers the following features belonging to each category. The URL lexical structure-based features include URL length, number of dots in the URL, number of hyphens in the domain, presence of security-sensitive words in the URL, length of the directories in the path of the URL, number of sub-directories in the path of the URL, presence or absence of IP in URL, count of tokens in the path, largest path token length, average path token length, length of the file, the total number of dots in the file, the total number of delimiters in the file, length of arguments, number of arguments, length of largest argument value, the maximum number of delimiters in arguments. The second category includes domain length, count of tokens in the domain, length of largest token in the domain, average domain token length, and suspicious top-level domain. The third category includes the age of the domain in months, the domain expiry age in months, the domain updating age in days, zip code of the address of the domain holder. The algorithms considered for a comparison study of the proposed phishing webpage detection are LR, NB, RF, DT, and $k$NN. The dataset considered is from the following sources: Detecting Malicious URL Machine Learning blob~\cite{master_dataset}, PhishTank~\cite{phishtank}, OpenPhish~\cite{openphish}, Majestic reports~\cite{majestic_reports}. The total dataset collected is $1,000,000$ which is split in the ratio of $7{:}3$ for training and testing the model. The performance metrics for the classifiers provided the best accuracy for the Naive-based and RF algorithms with accuracy values of $97.18\%$ and $98.03\%$ respectively. The AUC for the NB classifier is $99.1\%$.

Mohammad \textit{et al.}~\cite{alam2020phishing} proposed an ML approach for phishing webpage detection using a more informative feature set. The algorithm considered REF, Relief-F, IG and GR algorithms for extracting the best features from the dataset. The dataset includes $32$ features which are listed as a page index, long URL, short URL, presence of \texttt{@} symbol, redirecting \texttt{//}, prefix and suffix, subdomains, HTTPS, a domain registered length, favicon, non-standard port, HTTPS domain URL (TLS/ SSL), request URL, anchor URL, link in script tags, server form handler (SFH), info email, abnormal URL, website forwarding, status bar customization, disable right click, using a popup window, iframe redirection, age of the domain, DNS records, website traffic, Pagerank, Google index, links pointing to the page, class label. The PCA algorithm preserves the dataset information by reducing its variance to classify and identify the dataset components. The ML algorithms, RF and DT provide an accuracy of $96.96\%$ and $91.94\%$ respectively.

Rao \textit{et al.}~\cite{rao2020catchphish} proposed \textit{CatchPhish}, a phishing webpage detection approach that analyzes the suspicious URL string to mark it as phishing or legitimate. The feature extraction is obtained using two levels of extraction, the first one being the hand-crafted features and the second is TF-IDF algorithm-based feature extraction. The hand-crafted features contain $16$ host name-based features and $19$ full URL-based features. These two categories of hand-crafted features count the tokens often observed in phishing sites but rarely in legitimate sites, and also the top whitelisted brands that are targeted by the attackers. The second feature extraction algorithm used in TF-IDF computes the weights of the most frequent terms URL string. The next phase is the feature vectorization which takes an input vector of size $k$, which from the experiment analysis is obtained as $329,027$ which includes a vector of size $35$ from the hand-crafted feature and a vector of size $328,992$ obtained from the TF-IDF algorithm. The technique uses the ML algorithms: XG Boost, RF, LR, $k$NN, SVM, and DT which are trained on a dataset containing $42,220$ legitimate URLs from Common-crawl~\cite{common_crawl}, $43,189$ legitimate URLs from Alexa database~\cite{alexa_db} and $40,668$ phishing URLs from PhishTank~\cite{phishtank}. The five-fold data analysis on the benchmark database [phishtank $+$ Alexa, common-crawl $+$ PhishTank, a mixture of legitimate from common-crawl and Alexa $+$ PhishTank, legitimate from Yandex~\cite{yandex} $+$ PhishTank, legitimate from DMOZ~\cite{dmoz} $+$ phishtank] achieves an accuracy of $94.26\%$ for the RF algorithm when both the feature extraction algorithms were considered. The comparative analysis of the CatchPhish with \cite{sahingoz2019machine, marchal2014phishstorm} produced an accuracy: $98.25\%$, F1 score: $98.23\%$, precision: $98.04\%$ and sensitivity: $98.42\%$ for the RF algorithm.

\subsubsection{DL-based Techniques}
The DL approach is often used for visual analysis wherein the input is transformed to a vector of numbers called \textit{embedding}. In the DL approach for classifying the phishing URLs, the input URLs are transformed into a matrix where each row corresponds to a character.
\par
Le \textit{et al.}~\cite{le2018urlnet} proposed URLNet, which adopts CNN to process both the characters and words present in URL strings, aiming to acquire an embedded representation of URLs within a framework that is jointly optimized. This methodology enables the model to encompass diverse semantic information, surpassing the capabilities of previous models. Furthermore, it introduced enhanced word embeddings to address the challenge posed by an abundance of infrequently occurring words in the task. Through extensive experiments conducted on a substantial dataset, the results demonstrate a notable improvement in performance compared to the existing solutions.

\subsection{Webpage-based Approaches}
\label{sec:Webpage_Content-based_Approach}
Phishing webpage detection involves comparing the content of the input URL's webpage with a legitimate webpage. This method entails visiting the input URL's webpage and parsing it to obtain the Document Object Model (DOM), which provides attributes and values on the webpage. The attacker generates an input URL that appears identical to the legitimate webpage. The phishing site is designed to mimic the layout, logos, text fields, forms, font size, and images of a legitimate website. This approach can be further categorized into content similarity-based evaluation and ML-based approaches for similarity checks of webpages. Content-based approaches analyze features like term frequency, CSS similarity, hyperlink-based metrics, and image similarity scores to classify the suspicious input webpage in comparison with the legitimate page.

\noindent
\subsubsection{Content Similarity Techniques}
The webpage similarity is obtained from various features that are extracted from the webpage of the given URL compared against the legitimate webpage. The input URL webpage is visited to obtain the most frequent terms using the TF-IDF algorithm. The concatenation of the webpage domain name with the webpage's frequent terms is called lexical signature. The lexical signature is provided as an input query on the address bar of a browser to obtain a list of top $N$ URLs. The input webpages content such as HTML-based features like alarm window, restricted information, empty links, broken links, redirection, form content, brand name, and internal and external CSS features like color properties, font size, padding, margin, etc. are considered in obtaining the similarity score with the corresponding $N$ webpages in a sequential comparison test. This will yield a similarity score that can be used to classify the input webpage as phishing or legitimate at some threshold value. Figure~\ref{fig:list_heuristic_webpage_content_based} illustrates the algorithm employed for classifying a suspicious input webpage as legitimate or phishing. The algorithm starts by taking an input suspicious URL and then visiting the corresponding webpage. Subsequently, the webpage's source code is crawled to extract pertinent content-based features, in addition to performing URL analysis. The phishing data repository contains the URL and content patterns observed from previously flagged phishing webpages. By assessing the similarity of patterns matched in the suspicious webpage, the algorithm determines whether it should be classified as phishing or legitimate.


Zhang \textit{et al.}~\cite{zhang2007cantina} proposed CANTINA, a content-based approach for detecting phishing webpages. CANTINA uses the features like age of the domain, known images, presence of \texttt{@} and \texttt{-} in URL, suspicious links, IP address, dots in URL, and forms containing \texttt{<input>} tags with \texttt{credit card} and \texttt{password} terms. The algorithm works as for a given input URL, the webpage of the input URL is visited and the TF-IDF value of all the terms of the webpage is calculated. A lexical signature is generated that contains the top five frequent terms that are searched on Google as an input query. This helps in obtaining the top $n$ websites due to the query search. If the domain name of the input webpage goes with the domain name of the top $n$ results, then the input site is marked as legitimate otherwise, it is marked as phishing. The experiment analysis is carried out on phishing and legitimate URLs in four variable conditions: TF-IDF, TF-IDF $+$ domain, TF-IDF $+$ Zero results Meaning Phishing (ZMP) and TF-IDF $+$ domain $+$ ZMP. The test results show that TF-IDF $+$ domain $+$ ZMP achieves a TP of $97\%$ and an FP of $10\%$. The second experiment analysis is performed using weighted features heuristics along with TF-IDF $+$ domain $+$ ZMP that outperformed the anti-phishing tools SpoofGuard~\cite{teraguchi2004client} and Netcraft~\cite{netcraft} and obtained a TP of $89\%$.

\subsubsection{ML-based Techniques}
As seen in the case of URL-based evaluation, the ML algorithms help in extracting the features from a training dataset which leads to greater accuracy levels for classification. The same approach of using ML algorithms in the webpage content-based similarity evaluation techniques is leveraged. The ML algorithms can extract features from the HTML, CSS, Javascript and XML of the webpage source code and produce a feature vector to the ML classifier for performing the classification. Figure~\ref{fig:ml_url_content_based} shows the use of ML algorithms for better feature extraction to perform the webpage content analysis to find the legitimacy of the input webpage.


Ankit \textit{et al.}~\cite{jain2019machine} proposed a novel approach by inspecting only hyperlinks found in the HTML source code of the suspicious input page. The algorithm runs on the client side and considers only $12$ hyperlinks features for training the ML models. The $12$ hyperlink features considered are complete hyperlinks present in the source code of the webpage, no hyperlinks, internal/ external hyperlinks, null hyperlinks, internal/external CSS, internal/external re-directions, internal/external error, login form links in action fields, internal/ external favicon. These are the corresponding features for training the ML algorithms: SMO, NB, RF, SVM, AdaBoost, NN, C$4.5$ and LR. These ML algorithms are trained on the phishing URLs selected from PhishTank~\cite{phishtank} and legitimate URLs considered in combination with Alexa top websites~\cite{alexa_db}, Stuffgate's free online website analyzer~\cite{stuffgate} and a list of online payment service providers~\cite{payment_service_providers}. The algorithm's primary aim is to be language-independent and to detect phishing sites without considering any third-party-based services. The proposed algorithm outperforms \cite{pan2006anomaly, zhang2007cantina, garera2007framework, aburrous2010intelligent, whittaker2010large, xiang2011cantina+, he2011efficient, zhang2017two, el2017detection, montazer2015detection} in terms of search engine independence, language independence, zero-hour detection and third-party independence. The algorithms \cite{garera2007framework, zhang2017two} were helpful in detection, however, these two algorithms were third-party dependent. The proposed algorithm achieved the best performance metrics for LR algorithms with a TP rate of $98.39\%$, TN rate of $98.48\%$, precision of $98.8\%$, F1 score of $98.59\%$ and accuracy of $98.42\%$.

\subsubsection{Webpage Screenshot-based Techniques}
As observed in the webpage content-based approach, the textual content of the phishing webpage is compared with a legitimate webpage. Based on the similarity score and the domain name, the webpage is marked as phishing or legitimate. An approach that can rub parallel with the content analysis is the comparison of the webpage screenshot. In cases, where the text is replaced by the image, then the content-based analysis will not define an accurate similarity score. Hence, through webpage screenshots, it becomes easy to compare two exact-looking webpages and obtain the similarity score. In literature, there are several ways to obtain the similarity score between the two given webpages~\cite{hara2009visual, mao2017phishing, fu2006detecting, medvet2008visual, afroz2011phishzoo, lin2021phishpedia}.

Masanori \textit{et al.}~\cite{hara2009visual} proposed a visual similarity-based approach to detecting webpages that impersonate legitimate webpages. The algorithm takes in an input URL from PhishTank~\cite{phishtank} and accesses its webpage by visiting it and capturing its image using ImgSeek~\cite{jacobs1995fast}. The captured image of the input URL is compared with all the images one at a time from a database of images of legitimate webpages. The image database contains image screenshots marked with the domain name. If the image of the suspicious input webpage does not match the image of the legitimate webpage in the database then it is marked as an unknown site that can be a newly registered page and is added to the image database. Otherwise, the domain name of both pages is compared. If the domain name matches then the page is marked as legitimate otherwise phishing. The experimental analysis gave a TP of $80\%$ with an FP of $18\%$.

Mao \textit{et al.}~\cite{mao2017phishing} proposed an algorithm to quantify the suspiciousness of webpages relying on their visual appearance. The algorithm considers cascading style sheets (CSS) as a method to obtain the visual similarity score between the suspicious webpage and the legitimate webpage. Since the webpage elements do not contain similar influences, the proposed algorithm considers weighted page-component similarity. The dataset has $9,307$ phishing webpages collected from PhishTank that are targeting eBay, Paypal, Apple, and other popular websites out of which $6,192$ webpages are used for training and $3,115$ are used for obtaining the correctness of the model, and most targeted legitimate sites as a whitelist dataset. The CSS rules of both the suspicious page and the legitimate page are converted to the influence vector which is a normalized representation of the CSS rules. The influence vector includes two fields namely, property and declarations of values and selectors. The similarity between the influence vectors of both the suspicious webpage and the legitimate webpage is calculated and if the score is more than a threshold then, the webpage is marked as phishing else legitimate. The similarity score is obtained using the complexity score and match score.


The experimental analysis obtained a precision of $100\%$, recall of $97.92\%$, and F1 score of $98.95\%$. The comparison of the proposed algorithm with CANTINA~\cite{zhang2007cantina}, CANTINA+~\cite{xiang2011cantina+}, Corbetta \textit{et al.}~\cite{corbetta2014eyes}, proved that PhishAlarm obtained a higher precision and F1 score of $99\%$.

Fu \textit{et al.}~\cite{fu2006detecting} proposed an approach to compare two webpage screenshots to obtain the similarity score using Earth Mover's Distance (EMD) technique~\cite{rubner2000earth}. However, there is a possibility that the CSS styling of a few of the elements in the target legitimate webpage may vary over time, also, for dynamic webpages, the contents change frequently. Hence, the approach becomes computationally time-consuming with the comparison of the phishing webpage with a large number of reference webpages in the database.

Medvet \textit{et al.}~\cite{medvet2008visual} proposed a novel phishing webpage detection technique for visual similarity-based comparison of the suspicious input webpage with the legitimate webpage. The technique involved a comparison of the two webpages on a signature-based similarity value which is extracted using the visible text and image features. The algorithm retrieves the webpage of the input suspicious URL and finds the signature of the webpage. The signature of the suspicious input webpage is compared with the signature of the legitimate webpage. The signature for the suspicious URL webpage is extracted using the visible text features that include: text content, foreground color, background color, font size, font family, and visible image features that include: the area of the image in pixels, colour histogram, $2$D Haar wavelet transformation and image position. The comparison between the signature of the suspicious webpage and the legitimate page is obtained by comparing the signature score of the text $s^t$, the signature score of the image $s^i$ and the overall similarity score $s^o$. The value of the final similarity score is a sum of the $s^t$, $s^i$, and $s^o$ with coefficients. The comparison of the webpages is computed using pairwise comparison along each dimension. The experiment included phishing pages from PhishTank with $41$ positive pairs (phishing site, legitimate site) and $161$ negative pairs. The experimental results obtained a $0\%$ FP rate and $7.4\%$ FN rate.

Moreover, storing the top million legitimate webpages in a database is a cumbersome task. Hence, the latest approaches save only the webpage logos with the domain name. Afroz \textit{et al.}~\cite{afroz2011phishzoo} proposed a technique that compares the phishing webpage screenshot with the logos in the database. Comparing logos of a questionable website to those of established brands in a reference database allows for flexibility in webpage variations and designs. Furthermore, advancements in techniques like Scale-Invariant Feature Transform (SIFT)~\cite{burger2022scale} enable the comparison of images with variations in scale and orientation. Nonetheless, SIFT-based methods used by Afroz \textit{et al.}~\cite{afroz2011phishzoo} are not only computationally demanding, but they also lack precision.

The concern of high computations and lack of accuracy in the visual-based phishing webpage detection approaches are overcome by a phishing identification system called \textit{Phishpedia}~\cite{lin2021phishpedia}. It demonstrates superior performance compared to other identification approaches, boasting high accuracy in identifying phishing websites while maintaining low runtime overhead. Phishpedia uses Siamese Neural Network to recognize the brand. Siamese Neural Network helps to ease the task of comparing two images by forming their vectors. Also, Phishpedia offers explanatory annotations on webpage screenshots to aid in understanding phishing reports, making it highly practical. Evaluating Phishpedia using a six-month dataset of phishing URLs obtained from OpenPhish (Premium service), the assessment showcases the effectiveness and efficiency of the system. Additionally, within a 30-day period, Phishpedia successfully detected $1,704$ authentic phishing webpages, resulting in the collection of the largest dataset for evaluating phishing identification solutions, including information on phishing brands.

\subsubsection{DL-based Techniques on Webpage Screenshots}

\begin{figure*}[t]
    \centering
    \includegraphics[width=1.0\textwidth]{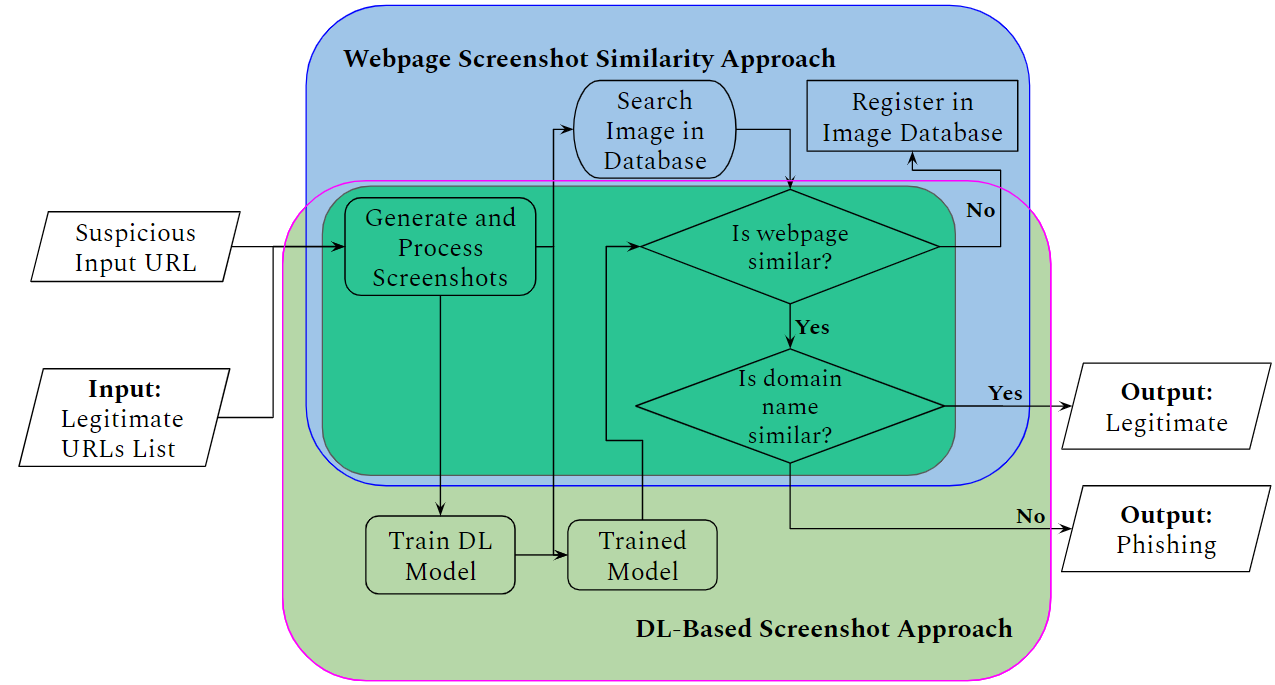}
    \caption{Webpage Screenshot Approaches (Similarity and DL-based)}
    \label{fig:webpage_screenshot_similarity_approaches}
\end{figure*}

Webpage screenshot-based phishing detection has been driven by the remarkable achievements of DL in the realm of Computer Vision. Figure~\ref{fig:webpage_screenshot_similarity_approaches} outlines a generic process for classifying a suspicious webpage as phishing. The algorithm in this process begins by considering a suspicious URL and then visiting the corresponding webpage to obtain the entire webpage screenshot. The suspicious screenshot is then processed to check for a match in a database containing images of legitimate webpages. If no similar screenshots are found, the webpage screenshot along with its URL is added to the image database. On the other hand, if the suspicious screenshot matches with one or more of the database screenshots, the algorithm proceeds to analyze the domain name of the suspicious URL against the domain name of the matched database screenshot. If the suspicious screenshot is similar to the image, but the domain names do not match, the suspicious URL is marked as phishing. Conversely, if the domain names match and the screenshots are similar, the webpage is classified as legitimate. 

Yi \textit{et al.}~\cite{yi2018web} conducts an analysis of the characteristics exhibited by phishing websites and proposes two types of features for detecting webpage-based phishing. These features include original features (URL features like count of special characters in URL and age of the domain) and interaction features (website interaction, and in-degree, and out-degree of URL). The study then proceeds to introduce Deep Belief Networks (DBN)~\cite{hua2015deep} as a means to identify phishing websites, explaining the detection model and algorithm employed by DBN. The authors train DBN using a small dataset to obtain suitable parameters for detection. Finally, the effectiveness of DBN is tested on a large dataset, resulting in $\sim 90\%$.

VisualPhishNet~\cite{abdelnabi2020visualphishnet} employs screenshots of secure websites from a designated reference list to train a Siamese neural network~\cite{chicco2021siamese}. During operation, the trained model evaluates the resemblance between a given screenshot of a webpage, which may be suspicious, and all the screenshots within the reference list. A webpage is classified as a phishing page if it exhibits significant similarity to any website in the reference list, despite having a distinct domain. This approach enables the detection of potential phishing pages by identifying visual similarities to known protected websites, providing an effective mechanism for distinguishing between legitimate and fraudulent web content.

Figure~\ref{fig:webpage_screenshot_similarity_approaches} presents the process of analyzing a suspicious URL based on its visual appearance using DL approaches on the webpage screenshot. The algorithm starts by considering a reference list of legitimate URLs, visiting each of them, and obtaining their screenshots, which are then stored in a database. Subsequently, DL algorithms are employed to analyze patterns in each of the images. When a suspicious URL is encountered, its corresponding webpage is visited to download the screenshot. The suspicious screenshot is then analyzed by the trained DL model to determine if it contains any patterns matching with the processed legitimate screenshots. If the suspicious screenshot does not match with any of the legitimate screenshots, it is marked as legitimate. However, if the suspicious screenshot matches with a legitimate screenshot, the algorithm proceeds to the next step of decision-making, which involves checking if the domain names match. If the domain name does not match, the suspicious URL is marked as phishing. Conversely, if the domain names match, the suspicious URL is marked as legitimate.


\subsection{Hybrid Approaches}
\label{sec:Hybrid_Approach}
The hybrid approach of phishing webpage detection combines all the existing detection techniques with the features from URL-based evaluation and content-based evaluation to form a feature vector that is provided to an ML algorithm for classification. The research works have proven that the hybrid approach produces a high accuracy level and also yields better performance results as compared to the previous two approaches (see, \ref{sec:URL-based_Approach}, \ref{sec:Webpage_Content-based_Approach}). In this approach, for a given input URL, the feature extractor extracts the effective features from the URL and stores them as a first vector, then the input URL webpage is visited to obtain frequent terms to form search queries and top $N$ matching results. The second vector is formed through the extracted features of the content-based similarities. These two vectors are then concatenated and provided in feature vectorization for the ML classifier. Since it considers both the URL and content features for prediction, it outperforms the previous two approaches (see, \ref{sec:URL-based_Approach}, \ref{sec:Webpage_Content-based_Approach}).

Srinivasa \textit{et al.}~\cite{rao2019detection} proposed a heuristic-based ML algorithm for finding the legitimacy of the input URL. The approach of the work initially parses the webpage of the suspicious input URL and extracts significant features from the webpage. The feature extraction phase includes three URL obfuscation features (dots in the hostname, URL with \texttt{@} symbol, lengthy URLs, using IP address, presence of HTTPS), three third-party-based features (age of domain, page rank, website in search engine results) and eight hyperlink-based features (frequency of domain in anchor links, frequency of domains in CSS links, image links and scripts links, common page detection ratio in website, common page detection ratio in footer, null links ratio in website, null links ratio in footer, presence of anchor links in HTML body, broken links ratio). The proposed work considers eight ML algorithms namely, RF, J$48$, LR, BN, MLP, SMO, AdaBoostM1, and SVM for comparison study among them to obtain the best classifier with the lowest FP rates. The algorithms consider $1,407$ legitimate URLs from Alexa and $2,119$ phishing URLs from PhishTank~\cite{phishtank} in the ratio of $75{:}25$ for training and testing the ML algorithms. The experimental analysis details the performance metric on all the ML algorithms used and RF achieves $99.3\%$ accuracy by considering third-party-based features. Furthermore, PCA-RF performs better than the oRF~\cite{menze2011oblique} algorithms and achieves an accuracy of $99.55\%$ and a precision of $99.45\%$.

Somesha \textit{et al.}~\cite{somesha2020efficient} proposed a DL model that can help in obtaining the legitimacy of the input URL. To achieve this, the authors considered three categories of features based on: URL, hyperlinks and third-party as a measure in the feature extraction phase for the given input URL. The feature selection phase considers the IG algorithm in determining the best performing and selective features out of $18$ features selected from the above-given feature categories that help in proving the legitimacy of input URLs. The IG algorithm produces the $10$ best features for the classification that includes three URL-based features (dots in hostname, URL length, presence of HTTPS), one third-party-based feature (page rank), and six hyperlinks-based features (presence of domain in anchor links, frequency of domain in image links, common page detection ratio, common page detection ratio in footer, presence of anchor tag in HTML body, broken links ratio). The DL model considered datasets from PhishTank~\cite{phishtank} and Alexa~\cite{alexa_db} with sizes $2,119$ and $1,407$ respectively. The DL algorithm is a DNN, LSTM, and CNN that are trained on the dataset provided and produced accuracy values of $99.52\%$, $99.57\%$, and $99.43\%$ respectively. In terms of accuracy ~\cite{somesha2020efficient} outperforms \cite{zhang2007cantina} by $89.18\%$ and~\cite{xiang2011cantina+} $99.13\%$.

\section{Discussion}
\label{sec:Discussion}
This section presents a comprehensive analysis of existing literature on phishing webpage detection solutions. The analysis encompasses discussions on the necessary inputs for phishing webpage detection, diverse repositories used for collecting phishing and legitimate samples, and a range of techniques employed, from traditional list-based approaches to the integration of ML and DL for zero-day phishing detection. 

\subsection{Inputs for Phishing Webpage Detection Approaches}
\label{sec:Inputs_Discussion}
In this section, we discuss the inputs considered by phishing webpage detection algorithms, which can be based on URLs, webpage content, or webpage screenshots. These inputs are necessary as they form the fundamental features relied upon by phishing webpage detection algorithms for classifying zero-day URLs as phishing or legitimate.

Regarding URL inputs, certain standards in the URL structure and the presence of prohibited special characters can mark a URL as suspicious, warranting further analysis by detection algorithms. Additionally, webpage content is also significant in assessing the legitimacy of a webpage. Features like anchors, scripts, link tags, and CSS rules are considered in this analysis.

Among the URL approach features, the presence of an IP address, HTTPS in the domain part of the URL, redirection, and the HTTPS protocol are commonly preferred. Similarly, the most utilized webpage content features include the presence of a form tag, the content of the action field in the form tag, and customized mouse hover. Visual features play a role as well, with analysis based on the webpage screenshot, logo, favicons, and the placement of tags on the webpage.

\subsection{Datasets}
\label{sec:comparison_based_on_dataset}
Phishing webpage detection solutions utilize phishing and legitimate samples from open-source repositories, which offer combinations of URL and/or webpage content, along with screenshots. These datasets are employed to address the phishing webpage detection problem, aiming to correctly classify zero-day phishing URLs or webpages. While most researchers rely on collecting phishing and legitimate inputs from these repositories, a few invest months in gathering the latest active phishing webpages to access all available features. Building an efficient model requires a diverse dataset. Table~\ref{tab:dataset} presents various phishing webpage detection approaches and their chosen repositories for collecting phishing and legitimate samples.

PhishTank~\cite{phishtank} emerges as the most preferred repository for collecting phishing webpages since it provides URLs and webpage screenshots. On the other hand, Alexa~\cite{alexa_db} is the favoured repository for collecting legitimate samples due to its comprehensive list of the top million domains.

\begin{table}[!h]
\caption{Dataset Repositories Considered by Research Works}
\label{tab:dataset}
\centering
\begin{tabular}{ll}
\toprule
\textbf{Repositories} & \textbf{Research Works} \\  \midrule
Alexa~\cite{alexa_db} & \cite{torroledo2018hunting, verma2015character, shirazi2018kn0w, rao2020catchphish, jain2018phish, rao2019detection, somesha2020efficient} \\
DMOZ~\cite{dmoz} & \cite{verma2015character, jain2018phish, rao2020catchphish} \\
Stuffgate~\cite{stuffgate} & \cite{jain2018phish} \\
Common Crawl~\cite{common_crawl} & \cite{rao2020catchphish} \\
Yandex~\cite{yandex} & \cite{rao2020catchphish} \\
Majestic~\cite{majestic_reports} & \cite{kumar2020phishing} \\
PhishTank~\cite{phishtank} & \cite{ abdelhamid2017phishing, jain2018phish, kumar2020phishing,  zhang2007cantina, hara2009visual, mao2017phishing, medvet2008visual, rao2019detection, somesha2020efficient} \\ & \quad \cite{verma2015character,rao2020catchphish, shirazi2018kn0w} \\
OpenPhish~\cite{openphish} & \cite{shirazi2018kn0w, kumar2020phishing, lin2021phishpedia} \\
UCI~\cite{UCI1} & \cite{sonmez2018phishing} \\
Millersmiles~\cite{millersmiles} & \cite{abdelhamid2017phishing} \\
\bottomrule
\end{tabular}
\end{table}

\subsection{Phishing Webpage Detection Methodologies}
\label{sec:Methodology_Discussion}
The phishing webpage detection techniques consider features based on URL, content, or visuals for finding the legitimacy of a zero-day URL. With novel phishing webpage detection methods, the FP rate is always high which means that the input URL is phishing, but is classified as legitimate by the observation, and thus, such URLs can easily be in use and continue extracting sensitive information from unsuspecting victims~\cite{amiri2014machine}. Moreover, these techniques also have low-accuracy values which does not characterize an efficient system. The list-based techniques cannot effectively identify all the phishing URLs in a feasible time which is sufficient for the attacker to do the damage and also it does not detect zero-day phishing URLs. Considering the visual-similarity-based approach also requires a large amount of memory to store the image database for comparison with the input webpage. Hence, lightweight techniques are required to reduce space complexity and computation cost~\cite{amiri2014machine}. Implementing such approaches using ML algorithms readily reduces human effort in terms of manual comparison and detection of URLs to add them to the blacklist. Moreover, it also helps in extracting effective features that are more suitable to make the classification. It is also observed that the model's performance improves if it considers all the heuristic features of URLs, content-based and styling-similarity.


\subsection{Performance Metrics}
\label{sec:comparison_based_on_performance_metrics}
The overall process leading to the final result involves feature extraction, feature selection, training and testing of classifiers on a specific dataset, and implementing the model to predict the output for zero-day input URL and classify it into the appropriate class by the classification model. Performance metrics are utilized to assess the effectiveness of the trained model. A confusion matrix illustrates the range of different class values based on the actual class and the predicted class, serving as a performance measurement for ML classification algorithms. The confusion matrix is used to calculate the model's accuracy, precision, recall, and F1 score using the terms TP, TN, FP, and FN.

Table~\ref{tab:comparisonofpapers} provides a statistical overview of various ML and DL-based phishing webpage detection solutions, listing the set of classifiers employed in each approach, along with the four performance metrics: accuracy, precision, recall, and F1 score. Researchers have conducted a comparative analysis to identify the best classifier that outperforms others for a given phishing webpage detection approach. The results depicted in Table~\ref{tab:comparisonofpapers} indicate that the RF classifier is predominantly favoured in most ML-based phishing webpage detection approaches due to its superior ability to accurately classify zero-day phishing URLs.

\begin{table}[tbhp]
    \caption{Comparison of Performance Metrics for Research Works}
    \label{tab:comparisonofpapers}
    \centering
    \begin{tabular}{p{1.2cm} p{1.8cm} p{0.35cm} p{0.35cm} p{0.35cm} p{0.5cm}}
    \toprule
       \multirow{5}*{Papers} & \multirow{5}*{Classifiers} & \multicolumn{4}{c}{\textbf{Performance Metrics (in \%)}} \\ \cline{3-6}
        ~ & ~ & \rotatebox{90}{Accuracy } & \rotatebox{90}{Precision} & \rotatebox{90}{Recall} & \rotatebox{90}{F1 score} \\ \midrule

        \cite{bahnsen2017classifying} & \begin{tabular}[c]{@{}l@{}}LSTM\\ RF\end{tabular} & \multicolumn{1}{c}{\begin{tabular}[c]{@{}c@{}}98.76\\ 93.47\end{tabular}} & \multicolumn{1}{c}{\begin{tabular}[c]{@{}c@{}}98.60\\ 93.64\end{tabular}} & \multicolumn{1}{c}{\begin{tabular}[c]{@{}c@{}}98.93\\ 93.28\end{tabular}} & \begin{tabular}[c]{@{}c@{}}98.76\\ 93.46\end{tabular} \\ \midrule 

        \cite{verma2015character} & \begin{tabular}[c]{@{}l@{}}PART\\ LR\\ J$48$\\ RF\\ SMO\\ NB\end{tabular} & \multicolumn{1}{c}{\begin{tabular}[c]{@{}c@{}}93.17\\ 90.45\\ 94.00\\ 95.24\\ 90.79\\ 77.34\end{tabular}} & \multicolumn{1}{c}{-} & \multicolumn{1}{c}{-} & - \\ \midrule 

        \cite{torroledo2018hunting} & - & \multicolumn{1}{c}{94.87} & \multicolumn{1}{c}{95.01} & \multicolumn{1}{c}{94.74} & 94.85 \\ \midrule 

        \cite{shirazi2018kn0w} & $k$NN & \multicolumn{1}{c}{98.80} & \multicolumn{1}{c}{-} & \multicolumn{1}{c}{-} & - \\ \midrule 

        \cite{rao2019detection} & \begin{tabular}[c]{@{}l@{}}RF\\ J$48$\\ LR\\ BN\\ MLP\\ SMO\\ AdaBoostM$1$\\ SVM\\ PCA-RF\end{tabular} & \multicolumn{1}{c}{\begin{tabular}[c]{@{}c@{}}99.31\\ 98.98\\ 95.22\\ 98.82\\ 95.12\\ 93.63\\ 97.18\\ 95.94\\ 99.55\end{tabular}} & \multicolumn{1}{c}{\begin{tabular}[c]{@{}c@{}}99.42\\ 99.30\\ 96.00\\ 98.83\\ 96.12\\ 94.76\\ 97.25\\ 96.09\\ 99.45\end{tabular}} & \multicolumn{1}{c}{-} & - \\ \midrule

        \cite{somesha2020efficient} & \begin{tabular}[c]{@{}l@{}}DNN\\ LSTM\\ CNN\end{tabular} & \multicolumn{1}{c}{\begin{tabular}[c]{@{}c@{}}99.52\\ 99.57\\ 99.43\end{tabular}} & \multicolumn{1}{c}{-} & \multicolumn{1}{c}{-} & - \\ \midrule

        \cite{sonmez2018phishing} & \begin{tabular}[c]{@{}l@{}}ELM\\ NB\\ SVM\end{tabular} & \multicolumn{1}{c}{\begin{tabular}[c]{@{}c@{}}95.34\\ 93.80\\ 92.98\end{tabular}} & \multicolumn{1}{c}{-} & \multicolumn{1}{c}{-} & - \\ \midrule

        \cite{rao2020catchphish} & \begin{tabular}[c]{@{}l@{}}XGBoost\\ RF\\ LR\\ $k$NN\\ SVM\\ DT\end{tabular} & \multicolumn{1}{c}{\begin{tabular}[c]{@{}c@{}}88.18\\ 94.26\\ 92.07\\ 84.92\\ 87.15\\ 91.71\end{tabular}} & \multicolumn{1}{c}{\begin{tabular}[c]{@{}c@{}}95.67\\ 98.59\\ 97.87\\ 88.15\\ 93.96\\ 94.62\end{tabular}} & \multicolumn{1}{c}{-} & \begin{tabular}[c]{@{}c@{}}91.64\\ 95.88\\ 94.36\\ 88.79\\ 90.83\\ 93.92\end{tabular} \\ \midrule

        \cite{rao2019jail} & - & \multicolumn{1}{c}{98.39} & \multicolumn{1}{c}{99.44} & \multicolumn{1}{c}{-} & 98.48 \\ \midrule 

        \cite{abdelhamid2017phishing} & \begin{tabular}[c]{@{}l@{}}C$4.5$\\ eDRI\\ RIDOR\\ OneRule\\ Conjunctive Rule\\ BN\\ SMO\\ AdaBoost\end{tabular} & \multicolumn{1}{c}{\begin{tabular}[c]{@{}c@{}}96.00\\ 94.00\\ 93.00\\ 89.00\\ 88.00\\ 93.00\\ 94.00\\ 92.00\end{tabular}} & \multicolumn{1}{c}{-} & \multicolumn{1}{c}{-} & - \\ \midrule

        \cite{alam2020phishing} & \begin{tabular}[c]{@{}l@{}}DT\\ PCA-RF\end{tabular} & \multicolumn{1}{c}{\begin{tabular}[c]{@{}c@{}}91.94\\ 96.96\end{tabular}} & \multicolumn{1}{c}{\begin{tabular}[c]{@{}c@{}}88.04\\ 96.89\end{tabular}} & \multicolumn{1}{c}{\begin{tabular}[c]{@{}c@{}}93.84\\ 42.16\end{tabular}} & \begin{tabular}[c]{@{}c@{}}90.57\\ 58.74\end{tabular} \\ \midrule

        \cite{jain2018phish} & \begin{tabular}[c]{@{}l@{}}NB\\ SVM\end{tabular} & \multicolumn{1}{c}{\begin{tabular}[c]{@{}c@{}}76.84\\ 91.28\end{tabular}} & \multicolumn{1}{c}{-} & \multicolumn{1}{c}{-} & - \\ \midrule

        \cite{alkawaz2021comprehensive} & - & \multicolumn{1}{c}{98.68} & \multicolumn{1}{c}{-} & \multicolumn{1}{c}{-} & - \\ \midrule

        \cite{abu2007comparison} & \begin{tabular}[c]{@{}l@{}}LR\\ CART\\ SVM\\ NN\\ BART\\ RF\end{tabular} & \multicolumn{1}{c}{-} & \multicolumn{1}{c}{\begin{tabular}[c]{@{}c@{}}95.11\\ 92.32\\ 92.08\\ 94.15\\ 94.18\\ 91.71\end{tabular}} & \multicolumn{1}{c}{\begin{tabular}[c]{@{}c@{}}82.96\\ 97.07\\ 82.74\\ 78.28\\ 81.08\\ 88.88\end{tabular}} & \begin{tabular}[c]{@{}c@{}}88.59\\ 89.59\\ 87.07\\ 85.45\\ 87.09\\ 90.24\end{tabular} \\ \midrule

        \cite{sharma2019phishalert} & - & \multicolumn{1}{c}{-} & \multicolumn{1}{c}{98.00} & \multicolumn{1}{c}{97.00} & 97.00 \\ \midrule

        \cite{jain2019machine} & \begin{tabular}[c]{@{}l@{}}SMO\\ NB\\ RF\\ SVM\\ AdaBoost$1$\\ NN\\ C$4.5$\\ LR\end{tabular} & \multicolumn{1}{c}{\begin{tabular}[c]{@{}c@{}}96.89\\ 95.79\\ 97.37\\ 91.47\\ 95.83\\ 97.25\\ 97.29\\ 98.42\end{tabular}} & \multicolumn{1}{c}{\begin{tabular}[c]{@{}c@{}}97.53\\ 96.67\\ 98.43\\ 92.19\\ 97.42\\ 97.41\\ 97.75\\ 98.80\end{tabular}} & \multicolumn{1}{c}{-} & \begin{tabular}[c]{@{}c@{}}97.22\\ 96.23\\ 97.63\\ 92.42\\ 96.24\\ 97.55\\ 97.58\\ 98.59\end{tabular} \\ \midrule

        \cite{zabihimayvan2019fuzzy} & - & \multicolumn{1}{c}{-} & \multicolumn{1}{c}{-} & \multicolumn{1}{c}{-} & 95.00 \\ \bottomrule
    \end{tabular}
\end{table}

\section{Evading Phishing Webpage Detection Models}
\label{sec:Current_Hurdles_in_Phishing_Detection}
In this section, we discuss the latest challenges in phishing URLs and webpage detection approaches. Attackers apply various tactics in initiating attacks that bypass the phishing webpage detection approaches to gather users' sensitive information. These tactics include \textit{hosting phishing webpages on compromised domains}, and \textit{using shortening services on phishing URLs}. Apart from these tactics, attackers also innovate in generating phishing webpages through AI and large language models (LLMs) and initiating \textit{adversarial attacks on phishing detection models}. The following paper summarizes the approaches to overcome such latest attack tactics in the phishing webpage detection domain.

\subsection{Phishing on Compromised Domains}
\label{sec:Phishing_on_Compromised_Domains}
Compromised domains, which are legally registered domains that are illicitly hacked for the purpose of hosting phishing webpages, represent a tactic employed by attackers to circumvent the detection mechanisms of search engine-based phishing detection techniques~\cite{rao2019jail}. The search engine-based techniques accept a search query that includes the domain name, keywords, plain text, and title. If the domain name specified in the search query appears within the retrieved list of website links, it is classified as legitimate else phishing. However, these techniques face a limitation when it comes to identifying whether a phishing webpage is hosted on a compromised domain. This limitation arises due to the compromised domain's domain age, which may have been indexed for an extended period, potentially leading to the false assumption of legitimacy~\cite{jain2018two}.

To address the complexities associated with detecting phishing webpages hosted on compromised domains, Rao \textit{et al.}~\cite{rao2019jail} introduced the \textit{Jail-Phish} algorithm. This approach considers an input URL, initiating a request to retrieve the webpage's source code, domain name, and title. Subsequently, a search query is constructed by appending the domain name with the title and submitting it to a search engine to obtain the top $n$ search results. If the domain from the query matches any of the domains in the search results, the content of the query webpage is compared with that of the matching search result content, generating a similarity score. If the similarity score exceeds a predefined threshold, the webpage is classified as \textit{legitimate} otherwise, it is classified as a \textit{phishing site hosted on a compromised domain}. However, it's important to note that for newly registered or less well-known legitimate domains, the search query may not yield any search results, potentially leading to an erroneous phishing classification. To mitigate this issue, the authors addressed this assumption by introducing a \textit{dynamic search query} approach. This modification significantly reduces the FP rate (legitimate as phishing) considerably.

Corona \textit{et al.}~\cite{corona2017deltaphish} proposed \textit{DeltaPhish} algorithm, which serves the purpose of assessing whether a given URL hosts a phishing webpage by conducting an analysis of both the HTML code and the visual differences between the input URL webpage and the corresponding website homepage. This algorithm initiates the process by requesting the input URL, to download the HTML source code and a screenshot of the webpage. Simultaneously, it requests the input domain homepage, downloading both its HTML code and a screenshot. The HTML code of both webpages (landing webpage and the homepage), undergoes a comprehensive analysis based on various HTML features. Likewise, the webpage screenshots are subjected to another feature extractor to compute visual features. The outputs from the HTML-based classifier and the Snapshot-based classifier yield dissimilarity scores that base the dissimilarities between the provided inputs. These dissimilarity scores are subsequently combined to generate an aggregated score. If the aggregated score is greater than or equal to $0$, the webpage is classified as phishing else legitimate.

\subsection{Detecting Phishing URLs via Shortening Services}
\label{sec:Detecting_Phishing_URLs_via_Shortening_Services}
Phishing URL detection models traditionally employ \textit{URL length} as a heuristic-based feature to classify legitimate and phishing URLs. This feature decides the class of a URL based on its character count. However, attackers are employing URL shortening services to shorten lengthy phishing URLs, resulting in what is commonly referred to as \textit{tiny URLs}~\cite{chhabra2011phi}. These tiny URLs obscure webpage details, such as the actual domain name and the path to phishing webpages embedded with malicious content and also can bypass rule-based feature selection methods~\cite{tang2021survey}.

To address the challenge of detecting URLs using shortening services like \texttt{bit.ly}, \texttt{goo.gl}, etc, Basnet \textit{et al.}~\cite{basnet2014learning} developed a Python library~\cite{pylongurl} utilizing the web service API provided by \texttt{longurl.org}. This library automatically expands URLs that have been subjected to shortening services. The approach takes into account a total of $333$ widely recognized shortening services. However, the algorithm encounters limitations in detecting tiny URLs either employed using new shortening services or by attackers creating their own shortening services, rendering traditional detection methods less effective in these cases.

Sameen \textit{et al.}~\cite{sameen2020phishhaven} also introduced \textit{URL Hit} as a foundational component within the PhishHaven algorithm, specifically designed for the detecting and classifying of tiny URLs. The PhishHaven algorithm is structured around four core components: \textit{URL Hit}, \textit{Feature Extractor}, \textit{Modeling}, and \textit{Decision Maker}. In the initial \textit{URL Hit} component, the input URL undergoes redirection to the PhishHaven browser plugin, which then initiates a request and response process. The response received may either be an expanded URL (in the case of a tiny URL) or the original URL. In the second component, referred to as \textit{Feature Extractor}, relevant URL-based features are extracted using regular expressions, focusing on the presence of seventeen specific characters within the response URL string. The resultant feature vector serves as input to ten distinct ML classifiers, constituting the third component. These classifiers operate independently to output prediction results. Finally, within the fourth component, the label for the input URL is determined through a \textit{voting} approach with a threshold of $67\%$. PhishHaven operates as a browser plugin, functioning as an intermediary between the client system and the server. Consequently, any input URL is initially directed to PhishHaven, which requests the server to retrieve a response webpage. This approach allows PhishHaven to analyze whether the input URL corresponds to a tiny URL or an expanded URL.

\subsection{Adversarial Attacks on Phishing Detection Models}
\label{sec:Adversarial_Attacks_on_Phishing_Detection_Models}
An adversarial attack is a technique to purposely develop an adversarial instance that can deceive a model into making mispredictions, reveal sensitive information or corrupt the model~\cite{adversarialAttackDefinition}. ML classifiers undergo training on datasets where experts meticulously curate pertinent features for optimum efficacy in detecting zero-day phishing attacks. However, these classifiers remain vulnerable to adversarial attacks, which endeavour to subvert their accuracy, potentially granting zero-day phishing attacks the ability to evade detection. This assessment typically relies on decision trees or rule sets~\cite{das2019sok}.

Attackers actively engage in data manipulation, aiming to induce false negatives~\cite{dalvi2004adversarial}. Attackers may introduce noisy data samples or craft entirely new attack instances by tampering with the characteristics of existing data. Noisy data samples result in classification models exhibiting reduced accuracy, necessitating heightened attacker effort. Conversely, feature manipulation presents a more insidious threat, enabling attackers to bypass established classifiers with relative ease~\cite{shirazi2019adversarial}.

\subsection{Phishing with AI: Threats \& Safeguards}
\label{sec:Phishing_with_AI:_Threats_&_Safeguards}
With the contemporary developments in large language models (LLMs) and ML, there has been a consequential surge in the technical aspect of phishing attacks. LLMs, exemplified by OpenAI's ChatGPT~\cite{openai_chatgpt} renowned for its prowess in code generation and textual analysis, have been harnessed for the creation of cyberattacks. While ChatGPT~\cite{openai_chatgpt} offers a multitude of advantages, such as crafting persuasive marketing content, scripting advertisements, generating code, and code troubleshooting, these very capabilities have enticed malicious actors to create large-scale social engineering attacks. It is important to note that ChatGPT has been designed to identify and reject prompts aimed at generating fraudulent, deceptive, or harmful content, in compliance with OpenAI's usage policies~\cite{linda_mini_tool}.

Roy \textit{et al.}~\cite{roy2023generating} delved into ChatGPT's potential to autonomously generate phishing websites. Their findings reveal that these generated phishing websites can convincingly mimic well-known brands and employ diverse evasion techniques to evade anti-phishing tools. The authors' motivation behind executing these attacks is to encompass a wide array of phishing techniques documented in the literature. By assessing ChatGPT's capability to generate such attacks, the authors aimed to highlight ChatGPT's potential impact on the security landscape and to raise awareness among security researchers, practitioners and users. In addition to generating conventional phishing attacks~\cite{alabdan2020phishing}, their study explores evasive attack vectors, including Captcha~\cite{kang2010captcha} and QR code-based~\cite{alnajjar2016trustqr} multi-stage attacks, attacks leveraging various evasive DOM features~\cite{liang2016cracking}, and many more.

Adversaries create phishing webpages to replicate legitimate webpages by incorporating malicious features designed to gather sensitive user information. Historically, attackers manually crafted phishing URLs; however, in recent years, there has been a shift towards autonomous URL generation using AI~\cite{vargas2016knowing}. This transformation aligns with the attackers' persistent pursuit of evading detection mechanisms. Acknowledging the strategic adaptability of attackers, Bahnsen \textit{et al.}~\cite{bahnsen2018deepphish} proposed \textit{DeepPhish} an algorithm built using LSTM networks. DeepPhish leverages knowledge acquired from prior successful attacks to autonomously create phishing URLs with the potential to bypass detection systems.

Sameen \textit{et al.}~\cite{sameen2020phishhaven} proposed an innovative real-time AI-generated phishing URL detection browser plugin named \textit{PhishHaven}, which represents a pioneering advancement in the field. PhishHaven is developed specifically for DeepPhish-generated phishing URLs. The algorithm deploys an ensemble ML model, leveraging lexical feature analysis for enhanced accuracy. PhishHaven achieved exceptional results in detecting AI-generated phishing URLs, with an accuracy rate of $98\%$, F1 score of $98\%$, TP rate of $97\%$, TN rate of $99.17\%$, and an impressively low FR rate of $0.8\%$.

In addition to creating phishing attacks using LLMs like ChatGPT, researchers have explored the potential of these models in automating the detection of phishing webpages. This advancement aims to enhance the efficiency of phishing webpage detection without requiring extensive fine-tuning of ML models dedicated to detection. Koide \textit{et al.}~\cite{koide2023detecting} introduced an innovative approach, a pioneering one in its category, for identifying phishing webpages by utilizing LLM (i.e., ChatGPT), optical character recognition (OCR), and web crawling techniques. This proposed approach crawls through an input webpage and generates various \textit{prompts}, namely \textit{plain text}, \textit{simplified HTML}, and \textit{URL}, which are passed to ChatGPT to determine if the input webpage is phishing or legitimate. The performance evaluation of the LLM is assessed using a dataset that includes evaluations conducted on both GPT-$3.5$ and GPT-$4$ models. Based on the experimental results of phishing webpage detection, GPT-$4$ outperformed GPT-$3.5$ with a precision, recall and FN rate reduction of $98.3\%$, $98.4\%$ and $11.7\%$ respectively.

\section{Open Issues with Potential Solutions}
\label{sec:Open_Issues}
In this section, we address the challenges and limitations faced by current phishing webpage detection approaches. ML classifiers tend to be biased towards larger datasets when trained on imbalanced data, resulting in lower accuracy for classifying zero-day phishing attacks. Models trained with limited datasets and considering correlated and less informative features struggle to accurately classify zero-day phishing URLs or webpages.

Visual similarity-based approaches, as proposed in \cite{zhang2007cantina, sharma2019phishalert, mao2017phishing, hara2009visual, medvet2008visual}, utilize non-ML-based methods to obtain similarity scores on text data. However, these approaches fail when textual data on the webpage is replaced with images containing the same textual content using the CSS background-image property.

Approaches discussed in \cite{rao2019detection, sonmez2018phishing} employ ML algorithms for classifying new URLs. Nevertheless, visual similarity-based approaches, described in \cite{xiang2011cantina+, zhang2007cantina}, are vulnerable to cross-site scripting (XSS). Mao \textit{et al.}~\cite{mao2017phishing} encounter limitations in dataset usage, impacting the model's efficiency on similarity scores. Approaches relying solely on textual data may fail to detect webpage cloning, and those heavily reliant on URL-based features face challenges when local DNS modifications are used to generate phishing URLs.

Current phishing webpage detection approaches encounter challenges when attempting to classify URLs generated through shortening services. Phishing URLs often are long so utilization of shortening services enables them to elude the phishing webpage detection solutions that rely on URL length as a discerning feature for distinguishing between phishing and legitimate URLs.

Visual similarity-based approaches should ensure the use of large datasets for comparing webpage content, which may necessitate storing webpage screenshots and result in high space consumption.

Based on an extensive review of various phishing webpage detection methodologies in the literature, we have identified several key challenges. In this section, we list the challenges observed in the existing phishing webpage detection solutions with their potential solutions.


\subsection{Datasets}
In the phishing webpage detection approaches, the collection of input samples is a pivotal task in analyzing the existing phishing and legitimate samples to draw insights into the heuristics for classifying zero-day phishing attacks. However, the dataset collection is an issue that needs to be resolved. Basically, we observe and discuss two issues in dataset collection, namely unbiased datasets and collection of diverse datasets as follows.

\textbf{Unbiased Datasets:} Classifier models leverage the provided training data to learn distinctive features for constructing decision trees or rule sets, with the most informative feature serving as the root node, enabling the classification of new suspicious inputs. In the case of a proportionate dataset, where the size of the phishing dataset is comparable to that of the legitimate dataset, the model can generate a well-aligned decision tree. However, when the dataset is imbalanced, the decision tree becomes skewed towards the dataset with a larger volume, implying a greater abundance of features for the model to learn from~\cite{das2019sok}. Consequently, the skewness property of the model classifies new suspicious URLs based on the features it has learned more extensively, thereby diminishing the accuracy rate of the model.

\textbf{Lack of Diverse Datasets:} When it comes to ML-based phishing detection, the model undergoes training using a provided set of features extracted from a dataset comprising both phishing and legitimate samples. However, it is crucial for the dataset to exhibit diversity. \textit{Diversity}, in this context, encompasses several aspects, such as including subpages and main domain landing pages as inputs, incorporating the use of short URLs in phishing webpages, and gathering samples that exhibit a variety of phishing characteristics. These characteristics may include malicious JavaScript links, iframes containing malicious forms, and images used as advertisements that contain links to phishing webpages. By encompassing this diverse range of samples, a large and varied dataset is made available for the classifier to train on. Consequently, the model becomes adept at accurately classifying new URLs into their appropriate categories.

\textbf{Potential Solutions:}
A potential solution to address the issue of an unbiased dataset involves employing a proportionate dataset comprising both categories, which is relatively straightforward. Nonetheless, the $k$-fold cross-validation approach, though computationally expensive, offers superior learning by assessing the model's performance through the subdivision of the dataset into $k$-folds. Moreover, regarding the issues pertaining to the datasets (unbiased and lack of diverse datasets), an effective solution involves creating a phishing webpage generation tool. This tool should function by taking a provided list of legitimate URLs and enhancing them with a \textit{randomly selected diverse set of content and visual attributes} typically associated with phishing. The URLs for the generated phishing webpages can be obtained by analyzing the characteristics of phishing URLs. To ensure accurate values for third-party-based features, it might not be necessary to host all the generated phishing webpages; instead, average notable values for these features can be assigned to the URLs based on studies. The output of this tool is a simulated phishing webpage with a dynamic nature, capable of operating on any given set of legitimate URLs as input and incorporating a stochastic process to include a subset of frequently observed phishing characteristics in the webpage content. By utilizing this innovative tool, it becomes possible to construct a dataset comprising an equal number of samples representing both legitimate and phishing webpages. As a result, the aforementioned challenges can be effectively mitigated, leading to a more balanced and diverse dataset for improved phishing webpage detection research.

    
\subsection{Feature set selection}
Several phishing webpage detection approaches strive to achieve the highest accuracy values for their proposed models by considering every available feature. While incorporating all features contributes to the model's ability to classify URLs, it suffers from impartiality and high computational costs. Alternatively, feature selection algorithms offer an alternative approach by identifying relevant features such as \textit{Information Gain (IG), Correlation-based Feature Selection (CFS), and Feature Ranking Score (FRS)}. In the context of phishing webpage detection, these approaches leverage URL-based and/or content-based features, or a combination of both, to design efficient models.

For URL-based feature selection, it becomes necessary to identify the most pertinent features belonging to categories such as address bar-based, domain-based, and certificate-based features. These selective features play a crucial role in accurately classifying new suspicious URLs as either phishing or legitimate. However, due to the presence of correlations among these features, considering all correlated features significantly increases the computational complexity of the algorithm. To mitigate this issue, feature selection algorithms strategically decide on the subset of features to consider, leading to improved performance and higher accuracy values.

Similarly, content-based features also exhibit correlations, necessitating the inclusion of feature selection algorithms to gather relevant features for constructing efficient anti-phishing algorithms. Moreover, it is advisable to independently run a set of classifiers for URL-based and content-based features, and subsequently identify the top features with higher information gain. By extracting these features from both the URL and content categories, the model's efficiency is enhanced, ultimately contributing to better performance.

\subsection{Enhancing Performance}
The performance of a model holds significant importance when it comes to effectively classifying zero-day phishing webpages. In the traditional ML approach for phishing webpage detection, a subset of features is extracted from a collected dataset during the training phase. Subsequently, the model's efficiency is tested on a small portion of the same dataset to evaluate its classification accuracy. However, if the model is not trained on a diverse dataset that includes representative samples of both phishing and legitimate instances and does not utilize an appropriate set of phishing features, it may erroneously classify a zero-day phishing webpage as legitimate.

To enhance the model's efficiency, a stacked model approach can be employed. This involves utilizing a series of ML classifiers at each stage of URL and webpage content classification. By implementing a stacked model, the performance of the model improves due to the construction of a well-structured decision tree or rule set, enabling accurate detection of new samples.
    
\subsection{Brand Prediction}
The process of forecasting potential target brands susceptible to phishing attacks involves an analysis of brand popularity based on factors such as profitability, production, sales, and fundamental details. Natural Language Processing (NLP) techniques can be employed for this purpose. Techniques such as web scraping, text extraction and parsing, information extraction, data aggregation, ranking, and report visualization prove valuable in extracting lists of top brands. These NLP techniques can be applied to websites like \textit{Forbes~\cite{forbes}, Interbrand~\cite{interbrand}, Fortune~\cite{fortune}, Bloomberg~\cite{bloomberg},} and \textit{BrandZ~\cite{brandz}}, which regularly publish relevant information about various brands. Consequently, the domain that consistently appears in the top lists generated through the NLP approach is likely to be targeted by attackers seeking to gather sensitive information.

\subsection{User Education}
Despite users' continuous internet surfing, it is evident that they may still fall victim to phishing webpages and unknowingly submit their credentials, even if they possess knowledge about phishing threats. In order to address user awareness regarding the authenticity of webpages,  the development of user education guides need to be developed. 
\textit{PhishGuru} and \textit{Anti-Phishing Phil}~\cite{kumaraguru2010teaching} are a few existing tools that serve this purpose of educating users. \textit{`PhishGuru'} uses email interventions, with the first intervention being more effective. It introduces text/graphic and comic strip interventions, with the comic strip proving the most effective. Anti-Phishing \textit{`Phil'} is a game teaching users to identify legitimate and phishing URLs, emphasizing the importance of examples, visual aids like comics, and user engagement for effective anti-phishing education and confidence in identifying phishing URLs.
However, users sometimes unknowingly fall for phishing attacks and thus, it also requires researchers to develop browser extensions that do the job of identifying the phishing webpages. The primary function of the browser extension would be to flag a webpage as safe if it is determined to be legitimate. Conversely, if a webpage exhibits characteristics indicative of phishing, it would be marked as malicious. Additionally, the browser extension would provide users with a comprehensive report, including a heatmap screenshot of the phishing webpage. This heatmap would highlight specific areas on the webpage that are deemed suspicious. Furthermore, the report would incorporate a list of the phishing features utilized within the webpage. This browser extension would serve as an on-the-spot educational tool for users, effectively alerting them when they encounter a potential phishing webpage. By providing users with a detailed list of phishing features and a visual heatmap report, they can be better equipped to recognize and avoid phishing threats in real-time.

\section{Related Surveys}
\label{sec:Contrast_with_Existing_Surveys}
In this section, we examine several review papers focused on phishing webpages detection. The selection of these review papers is guided by their publication records, citation counts, and an in-depth analysis of the phishing webpage detection domain. These selected review papers encompass works published between $2010$ and $2023$, each offering insights into the evolution of phishing webpage detection approaches over the years, including challenges encountered and the proposed solutions to address them.

Table~\ref{tab:Contrast_with_Existing_Survey_Papers} presents a subset of these selected review papers that are represented based on the discussion of various detection approaches, and research gaps provided by authors in respective papers with some ways to overcome those challenges. Each entry in Table~\ref{tab:Contrast_with_Existing_Survey_Papers} is accompanied by one of three labels: \textit{discussed} (indicated by \fullcirc[0.25em]), \textit{highlighted but not described with its potential solution} (marked as \halfcirc[0.3em]), or \textit{not mentioned in the paper} (left empty in the entry). Furthermore, Table~\ref{tab:Contrast_with_Existing_Survey_Papers} lists out a set of attributes over which these selected review papers are contrasted against our work. We put forth potential solutions to address these contrasting attributes in Section~\ref{sec:Open_Issues}. By adopting this comprehensive approach, our study makes a substantial contribution to the current body of knowledge and sheds light on emerging directions in the field of phishing webpage detection research.

\begin{table}[t]
    \caption{\centering Contrast with Existing Survey Papers}
    \label{tab:Contrast_with_Existing_Survey_Papers}
    \centering
    \begin{tabular}{p{1.4cm} p{0.18cm} p{0.18cm} p{0.18cm} p{0.18cm} p{0.18cm} p{0.18cm} p{0.18cm} p{0.18cm} p{0.18cm} p{0.18cm} p{0.18cm} p{0.18cm}}
    \toprule
       \multirow{20}{*}{\rotatebox{90}{\textbf{Papers}}} & \multicolumn{12}{c}{\textbf{Attributes}} \\ \cline{2-13}
        ~ & \rotatebox{90}{Feature Description} & \rotatebox{90}{Feature Selection Algorithm} & \rotatebox{90}{Performance Improvement Techniques}  & \rotatebox{90}{Shortening Service URLs Detection} & \rotatebox{90}{DL Approaches} & \rotatebox{90}{Labelled Dataset} & \rotatebox{90}{Dataset Sample Ratio} & \rotatebox{90}{Alternate for Dataset Collection} & \rotatebox{90}{Adversarial Attacks on Phishing Models} & \rotatebox{90}{Phishing Sites Hosted on Compromised Domains } & \rotatebox{90}{Brand Prediction} & \rotatebox{90}{LLM-generated Webpage Detection}  \\ \midrule
        
        \cite{vijayalakshmi2020web} & \fullcirc[0.35em] &  & \fullcirc[0.35em] &  &  &  &  &  &  &  &  & \\

         \cite{patil2019methodical} & \fullcirc[0.35em] & \fullcirc[0.35em] & \fullcirc[0.35em] &  & \halfcirc[0.35em] &  &  &  &  &  &  & \\

         \cite{do2022deep} & \fullcirc[0.35em] & \halfcirc[0.35em] & \fullcirc[0.35em] &  & \fullcirc[0.35em] & \fullcirc[0.35em] &  &  & \halfcirc[0.35em] &  &  & \\
        
        \cite{hannousse2021towards} & \fullcirc[0.35em] & \fullcirc[0.35em] & \fullcirc[0.35em] & \halfcirc[0.35em] &  &  & \fullcirc[0.35em] & \fullcirc[0.35em] &  &  &  & \\
        
        \cite{zieni2023phishing} & \fullcirc[0.35em] & \fullcirc[0.35em] & \fullcirc[0.35em] & \halfcirc[0.35em] & \halfcirc[0.35em] & & \fullcirc[0.35em] & & \halfcirc[0.35em] & \halfcirc[0.35em] & & \\
        
        \cite{catal2022applications} & \fullcirc[0.35em] & \fullcirc[0.35em] & \fullcirc[0.35em] & \halfcirc[0.35em] & \fullcirc[0.35em] & \fullcirc[0.35em] &  & \fullcirc[0.35em] &  &  &  & \\
        
        \cite{sahoo2017malicious} & \fullcirc[0.35em] & \fullcirc[0.35em] & \fullcirc[0.35em] & \halfcirc[0.35em] & \fullcirc[0.35em] & \fullcirc[0.35em] &  &  & \fullcirc[0.35em] & \halfcirc[0.35em] &  & \\
        
        \cite{jain2022survey} & \fullcirc[0.35em] & \fullcirc[0.35em] & \halfcirc[0.35em] & \halfcirc[0.35em] &  &  &  &  & \halfcirc[0.35em] & \fullcirc[0.35em] & \fullcirc[0.35em] & \\
        
        \cite{das2019sok} & \fullcirc[0.35em] & \fullcirc[0.35em] & \fullcirc[0.35em] & \halfcirc[0.35em] &  & \fullcirc[0.35em] & \fullcirc[0.35em] & \fullcirc[0.35em] &  &  & \fullcirc[0.35em] & \\
        
        \textbf{Our Work} & \fullcirc[0.35em] & \fullcirc[0.35em] & \fullcirc[0.35em] & \fullcirc[0.35em] & \fullcirc[0.35em] & \fullcirc[0.35em] & \fullcirc[0.35em] & \fullcirc[0.35em] & \fullcirc[0.35em] & \fullcirc[0.35em] & \fullcirc[0.35em] & \fullcirc[0.35em]
         \\ \bottomrule
    \end{tabular}
\end{table}

Zieni \textit{et al.}~\cite{zieni2023phishing} summarizes various phishing website detection approaches, categorizing them into three approaches: list-based, similarity-based, and ML-based and also provides issues pertaining to each approach. The list-based approach lacks the ability to detect zero-day attacks, making it vulnerable to emerging threats. The similarity-based approach analyzes both the webpage content and screenshots. However, the approach faces challenges when dealing with screenshot analysis when the text content on the webpage is replaced by the corresponding image using the CSS \texttt{background-image} property. The ML-based approach is effective in extracting features from URLs, and content, and visual-based for classifying webpages. Although powerful, it necessitates diverse datasets with an equal proportion of samples. The authors emphasize analysing the issues in developing robust and efficient phishing webpage detection systems.

Jain \textit{et al.}~\cite{jain2022survey} studied various phishing webpage detection approaches and explored attack techniques. The phishing webpage detection taxonomy includes URL-based, lookup-based, search engine-based, HTML-based, security-based, visual-based, and website traffic-based features. The paper lists the challenges in ML-based phishing detection approaches like language dependency and accuracy issues with compromised domains. Additionally, the study highlights that the search engine-based approaches are not suitable to combat DNS poisoning effectively, whereas the visual-based approach is effective but has a low response time. It provides valuable insights into phishing webpage detection approaches and their strengths and weaknesses.

Catal \textit{et al.}~\cite{catal2022applications} reviewed ML approaches for phishing webpage detection, consisting of supervised, unsupervised, semi-supervised, and reinforcement learning methods. They discussed feature extraction in these approaches and listed effective feature selection algorithms. The study focuses on using diverse datasets, feature selection and performance metrics to develop effective phishing webpage detection solutions. Furthermore, the research offers insights into the future scope of ML and DL approaches in phishing webpage detection while also providing a detailed examination of the limitations observed in existing approaches. These limitations include extended training times on biased datasets, the challenge of obtaining sufficient labelled datasets, detecting tiny URLs, the need for feature selection algorithms, and difficulties in detecting embedded objects within webpages. By acknowledging these limitations, the study contributes to the advancement of more robust and effective phishing webpage detection solutions.

Hannousse \textit{et al.}~\cite{hannousse2021towards} reviewed ML-based phishing webpage approaches, focusing on creating a diverse and unbiased benchmark dataset. The study conducted experiments by considering various features related to URLs, website content, and third-party sources. Relevant features were selected using filters like \textit{Chi-Square}, \textit{Pearson's correlation}, and \textit{IG} algorithms. The results identified the best algorithms within each feature set, concluding that a hybrid ML approach is most effective for improving the model's performance for zero-day phishing attack detection. Nevertheless, the study also identified certain limitations. The proportion of legitimate and phishing samples within the dataset was found to significantly impact the classifier's performance. Additionally, it was noted that failing to shuffle the dataset could lead to a serial correlation between the samples, potentially influencing the results. Being aware of these limitations provides valuable insights for future research in phishing webpage detection and classification.

Sahoo \textit{et al.}~\cite{sahoo2017malicious} conducted a comprehensive study on Malicious URL detection utilizing ML techniques. The paper offers a detailed explanation of the feature extraction categories, encompassing URL-based, webpage content-based, host name-based, third-party-based, and visual-based features. While the ML approach proves valuable for phishing URL detection, the authors also highlight certain issues that deserve attention. These issues mainly revolve around tiny URLs, challenges in dataset collection, and obtaining accurately labelled datasets. Furthermore, the study acknowledges several limitations in the context of the ML approach for phishing URL detection. These limitations include the difficulty in acquiring sufficiently labelled datasets, challenges in collecting efficient features tailored to the specific context of phishing webpage detection, and the susceptibility of the models to adversarial attacks. Being aware of these limitations will pave the way for further advancements and improvements in the field of malicious URL detection using ML methods.

Das \textit{et al.}~\cite{das2019sok} conducted an extensive review focusing on ML approaches for detecting phishing websites. The main objective of the authors is to emphasize the importance of adapting security measures in combating phishing attacks. The study thoroughly analyzes the various stages of phishing attacks and their associated challenges. In the set-up stage, challenges such as the lack of sufficient datasets and base-rate fallacy are addressed. During the feature extraction and ranking process, issues related to the time-scale of an attacker and false alarms are explored. Learning techniques face challenges with active attackers and the time-scale of attacks. The evaluation stage tackles concerns like base-rate fallacy, the cost of misclassification, and dealing with unbalanced datasets. Additionally, the insights section delves into issues like active attackers and the poisoning of public data sources. The research also identifies several gaps in the current landscape of phishing webpage detection. These gaps include the use of imbalanced datasets, improper metrics for evaluating classifiers, unreported training and testing times, and the lack of comprehensive generalization studies. Addressing these gaps is crucial in improving the effectiveness and efficiency of ML-based phishing website detection systems.

Vijayalakshmi \textit{et al.}~\cite{vijayalakshmi2020web} conducted a review of the ML approach for detecting phishing webpages and categorized the approaches into three main groups: web address-based (including list-based, heuristic rule-based, and learning-based methods), webpage content/similarity-based (comprising rule-based and ML-based techniques), and hybrid approaches. The study concludes that heuristic-based approaches demonstrate superior performance compared to other methods. However, the authors also identified certain limitations in the existing approaches. One of the limitations is the absence of feature selection in some methods, which could affect the efficiency of the classification process. Additionally, when dealing with non-English languages, these approaches may exhibit a high FP rate. In content-based detection, challenges arise in detecting text replaced with images, and setting appropriate threshold values for similarity rates of webpages proves to be difficult. Being aware of these limitations can guide further research and improvement in phishing webpage detection using ML techniques.

Do \textit{et al.}~\cite{do2022deep} reviewed phishing email and webpage detection approaches, categorizing them based on lists, heuristics, visual-similarity, ML, DL, and hybrid approaches. The paper concludes that the hybrid approach outperforms others, particularly in detecting zero-day phishing attacks. The paper primarily focuses on applying DL in cybersecurity, covering the detection of various attacks like phishing, malware, intrusion and spam. The taxonomy contains five categories for DL techniques: discriminative (supervised), generative (unsupervised), hybrid, ensemble, and reinforcement learning. Furthermore, it lists challenges related to feature selection, zero-day attack detection, dataset diversity, computational constraints, parameter tuning, metric selection for imbalanced datasets, interpretability of DL algorithms, batch learning, training time, and handling large datasets. The authors recommend that DL algorithms can address many of these challenges based on the selection of diverse datasets, features and relevant algorithms.

Alkawaz \textit{et al.}~\cite{alkawaz2021comprehensive} conducted an extensive review focusing on the ML approach for detecting phishing webpages. The authors provide a detailed examination of various aspects, including the taxonomy of phishing webpage detection, a comparison of different content-based phishing webpage detection methods, and a performance evaluation of PhishAlert, CANTINA, and CANTINA+ based on precision, recall, and F1 score metrics. The paper concludes that a hybrid approach proves to be more effective for phishing webpage detection. Moreover, the study highlights certain limitations. One such limitation is that the hybrid approach exhibits slower performance when conducting visual analysis on webpages for phishing detection. Additionally, phishing URLs that use HTTPS may require more time before they are flagged as phishing by secure browsers.

Singh \textit{et al.}~\cite{singh2020phishing} presents an extensive review of the ML-based approaches for detecting phishing webpages, encompassing heuristic, Fuzzy, ML, Image, Blacklist, and CANTINA methods. The authors delve into feature sets such as URL-based, source code-based, and image-based approaches. Additionally, the paper discusses various protection techniques employed to counter phishing attacks. However, the list-based approach fails to detect zero-day attacks, leaving it vulnerable to emerging threats. Moreover, client-side tools have their restrictions, being limited to examining phishing URLs and may not provide a comprehensive defence against more sophisticated phishing tactics.

Basit \textit{et al.}~\cite{basit2021comprehensive} surveyed ML, DL, scenario, and hybrid-based approaches. The study highlighted the superior performance of the RF classifier in ML approaches and explored various DL algorithms for feature extraction and pattern recognition. Scenario-based approaches focused on logos, email content, and attachments, while hybrid methods combined ML algorithms for improved performance. The paper concluded by emphasizing the effectiveness of ML algorithms like C$4.5$, XGB, $k$NN, SVM, and LR when used with feature extraction, achieving an accuracy of $99.2\%$. However, it acknowledged limitations, including low accuracy in existing phishing webpage detection tools, a high FP rate in hybrid-based approaches, and the computational efficiency of browser plug-ins for phishing webpage detection.

Patil \textit{et al.}~\cite{patil2019methodical} conducted a systematic review of phishing webpage detection approaches, categorizing them into list, rule (heuristics), content, ML, and hybrid-based techniques. The paper identifies limitations in existing anti-phishing tools, including Google Safe Browsing's reliance on lists and the occasional failure of tools like Avast~\cite{avast} and Quick Heal~\cite{quickheal} to detect zero-day attacks. To address these issues, the authors proposed enhancements, stressing the importance of diverse datasets for ML model accuracy and updated heuristics for improved performance. Feature selection algorithms are also highlighted to reduce training duration by selecting informative and less correlated features. The study found that heuristics and hybrid-based approaches outperformed others and suggested that neural networks could enhance the performance of phishing webpage detection approaches.

Prachit \textit{et al.}~\cite{prachit2020survey} undertook a comprehensive survey focusing on the detection of phishing webpages, wherein they classified URLs into three distinct categories: \textit{benign}, \textit{malware}, and \textit{spam}. The paper primarily presents a summary of multiple research papers, with a particular emphasis on approaches based on ML and DL. The summarized papers highlighted various features, encompassing URL characteristics, HTML attributes, CSS properties, Javascript behaviours, and third-party-related components.

Rana \textit{et al.}~\cite{alabdan2020phishing} conducted an extensive survey of phishing webpagedetection approaches, examining phishing mediums (Internet, SMS/MMS, voice-based) and vectors (email, fax, social media, websites, smishing, vishing). They categorized these approaches into two groups: \textit{traditional non-computerized anti-phishing techniques} involving legal actions and user education, and \textit{technical anti-phishing techniques} encompassing list-based, heuristics, ML, and visual similarity methods. Additionally, the paper enumerates the challenges encountered by phishing webpage detection models, such as the use of SSL for phishing webpages, the installation of malware on mobile devices, and the difficulty in pinpointing the source of a breach, given the increasing persistence of attackers.

Rahmad \textit{et al.}~\cite{abdillah2022phishing} conducted a systematic literature review on phishing webpage detection, emphasizing datasets, phishing attack techniques, and performance metrics. The paper provides insights into various attack sources for gathering user's sensitive information, through various means such as suspicious emails, browser pop-ups, SMS, and malicious tweets based on which the authors categorized the classification techniques into emails, URLs, and websites. Also, the use of ML classifiers in phishing webpage detection involves comparative study on various classifies out of which RF, SVM, and LR are used in around $39\%$ of the research works. Additionally, it discussed performance metrics used for comparing the most effective model for phishing webpage detection out of which the most used metrics are accuracy, TP rate, F1 score and precision. The authors identified open issues, including extending phishing webpage detection techniques to different languages beyond English, improving expert-based feature classification in suspicious emails, and defining standards in performance metrics values in phishing webpage detection solutions.

Safi \textit{et al.}~\cite{safi2023systematic} conducted a systematic review on phishing webpage detection approaches, and categorising them based on lists, heuristics, ML, DL, and visual similarity-based techniques. The paper meticulously scrutinizes the existing literature pertaining to these techniques, providing insights into the datasets used, the algorithms/classifiers utilized, key findings in the papers, and finally, the limitations as perceived by the authors. Drawing from their analysis of the collected papers, the authors observed that the most frequently used repository for gathering phishing URLs is PhishTank, while Alexa is the preferred choice for collecting legitimate URLs. Furthermore, their analysis revealed that the RF classifier is the most commonly employed ML classifier for phishing webpage detection.

Rastogi \textit{et al.}~\cite{rastogi2021survey} proposed an efficient phishing webpage detection model focusing on the fundamentals of ML-supervised classifiers. The author's perspective on the fact that the existing approaches are meant to minimize the impact of phishing attacks given the evolving and sophisticated tactics employed by attackers to circumvent current defences. In the study, four ML classifiers namely: KNN, Kernel-SVM, RF, and DT are used to create an efficient model by partitioning the dataset into training inputs, training outputs, testing inputs, and testing outputs.

Aung \textit{et al.}~\cite{aung2019survey} conducted a systematic literature review that centred on URL-based phishing detection approaches. They categorized existing phishing webpage detection approaches based on lists, content, visual similarity, and URL techniques. The paper provides a list of sixteen most frequently used URL-based features along with the most used ML classifiers including NB, SVM, RF, CNN, LR, LSTM, and DT. In summary, the authors conclude their review by highlighting several open issues in the field. They note that using imbalanced datasets can introduce bias into classification results, emphasizing that oversampling the minority class can enhance model effectiveness. Additionally, they necessitate collecting diverse datasets. For example, Alexa~\cite{alexa_db} provides top-ranked URLs of the index pages, however, repositories like PhishTank~\cite{phishtank} and OpenPhish~\cite{openphish} offer a diverse set of URLs. This diversity is vital because certain features, such as URL length, may introduce bias when applied to the Alexa dataset, but not when considering variable-length URLs collected from PhishTank and OpenPhish.

Khnoji \textit{et al.}~\cite{khonji2013phishing} conducted a systematic literature review by categorizing the phishing webpage detection approaches in measure of two sub-categories namely, user education and software detection techniques. The paper examines the human element involved, which includes phishing victims, user-phishing interaction models, service policies, passive and active warnings, and educational notices. In addition, it offers a comprehensive overview of various phishing webpage detection approaches, categorised based on lists, heuristics, visual-similarity analysis, and ML techniques. Moreover, the paper also highlights the challenges and limitations associated with these techniques. These encompass user education challenges, the complexities of classifying web pages rendered with natural language processing in multiple languages, and the limitations of offensive defence strategies and attackers' persistent efforts to deceive users and bypass the existing models.

Zuhair \textit{et al.}~\cite{zuhair2016feature} addresses the evolving challenges posed by intricate phishing webpages and the need for efficient phishing webpage detection approaches. While several detection methods exist, they often fall short due to the inadequate selection of features. The paper emphasizes the significance of selecting highly contributing features over less relevant ones to enhance detection system efficiency. It categorizes feature selection methods into two main categories: filter and embedded with classifiers. Filter selection includes feature weighting techniques like IG~\cite{pedregosa2011scikit}, CFS~\cite{michalak2006correlation}, and PCA~\cite{abdi2010principal}. Embedded methods encompass wrapper feature selection. The study underscores the importance of feature selection algorithms in identifying the most predictive feature subsets from diverse feature sets. It highlights key aspects of feature selection, such as assessing the contribution and quality of selected features, dealing with large and imbalanced datasets, addressing the heterogeneity of feature values, ensuring resilience in feature selection outcomes, managing relevance and redundancy, handling computational costs, and optimizing classification performance.

The selected review papers offered valuable insights into the various phishing webpage detection approaches by even listing out the challenges observed during the study. However, our research also uncovered new challenges within the phishing webpage detection domain. Additionally, alongside our analysis of review papers, we conducted an in-depth examination of the most recent technical papers in the domain. These technical papers specifically explored ML and DL approaches for phishing webpage detection. By combining insights drawn from existing literature with the latest developments in phishing webpage detection solutions, we successfully pinpointed several unresolved challenges and introduced potential solutions, thereby enriching the novelty of our study. Our study examines various surveys and identifies several unresolved challenges. These challenges include the absence of a diverse dataset, handling imbalanced datasets, dealing with correlated and less informative features during ML model training, finding solutions for collecting diverse datasets, creating models capable of accurately classifying zero-day phishing webpages and developing methods to educate users about phishing attacks. 

\section{Conclusion}
\label{sec:Conclusion_and_Future_Work}
In this survey, we categorized the existing approaches for phishing webpage detection based on the URL, webpage (content and screenshots), and hybrid approaches. Based on the study, the RF classifier outperforms other ML classifiers by offering high accuracy and a low FP rate. Moreover, the hybrid-based approach outperforms other approaches in providing high performance for zero-day phishing detection. Based on the survey, we list out several open issues in the phishing webpage detection approaches based on the dataset used, feature selection algorithms, performance metrics, and the latest challenges faced by phishing webpage detection models. We also provide potential solutions addressing a few of the open issues. Additionally, we provide steps for user education in the area of phishing.
Finally, we recommend collecting proportionate, diverse datasets in ML-based approaches and training them on features selected using feature selection algorithms. Also, the accuracy of the phishing webpage detection techniques can be improved by stacking ML classifiers. Moreover, the collection of dataset used by a research proposal should be saved in a repository which makes researchers comfortable for their future proposals to consider the same dataset for a meaningful comparative analysis.

\bibliographystyle{IEEEtran}
\balance
\bibliography{references}

\begin{thebibliography}{100}
\providecommand{\url}[1]{#1}
\csname url@samestyle\endcsname
\providecommand{\newblock}{\relax}
\providecommand{\bibinfo}[2]{#2}
\providecommand{\BIBentrySTDinterwordspacing}{\spaceskip=0pt\relax}
\providecommand{\BIBentryALTinterwordstretchfactor}{4}
\providecommand{\BIBentryALTinterwordspacing}{\spaceskip=\fontdimen2\font plus
\BIBentryALTinterwordstretchfactor\fontdimen3\font minus
  \fontdimen4\font\relax}
\providecommand{\BIBforeignlanguage}[2]{{%
\expandafter\ifx\csname l@#1\endcsname\relax
\typeout{** WARNING: IEEEtran.bst: No hyphenation pattern has been}%
\typeout{** loaded for the language `#1'. Using the pattern for}%
\typeout{** the default language instead.}%
\else
\language=\csname l@#1\endcsname
\fi
#2}}
\providecommand{\BIBdecl}{\relax}
\BIBdecl

\bibitem{ituStatistics}
``{Statistics, ITU},'' 2023,
  \url{https://www.itu.int/en/ITU-/Statistics/Pages/stat/default.aspx}
  [Accessed: October 14th, 2023].

\bibitem{itgovernanceBiggestPhishing}
L.~Irwin, ``{The 5 Biggest Phishing Scams of All Time - IT Governance Blog En,
  IT Governance European Blog},'' 2022,
  \url{https://www.itgovernance.eu/blog/en/the-5-biggest-phishing-scams-of-all-time}
  [Accessed: October 14th, 2023].

\bibitem{cisaWhatCybersecurity}
C.~. I. S.~A. (CISA), ``{What is Cybersecurity?}'' 2021,
  \url{https://www.cisa.gov/uscert/ncas/tips/ST04-001} [Accessed: October 14th,
  2023].

\bibitem{frauenstein2020susceptibility}
E.~D. Frauenstein and S.~Flowerday, ``{Susceptibility to phishing on social
  network sites: A personality information processing model},'' \emph{Computers
  \& security, Elsevier}, vol.~94, p. 101862, 2020.

\bibitem{apwg_report_2018}
APWG, ``{APWG Report},'' 2018,
  \url{https://docs.apwg.org//reports/apwg_trends_report_q1_2018.pdf}
  [Accessed: October 14th, 2023].

\bibitem{APWG_REPORT_4_2022}
------, ``{Phishing Activity Trends Report},'' 2022,
  \url{https://docs.apwg.org/reports/apwg_trends_report_q4_2022.pdf} [Accessed:
  May 13th, 2023].

\bibitem{cao2008anti}
Y.~Cao, W.~Han, and Y.~Le, ``{Anti-phishing based on automated individual
  white-list},'' in \emph{Proceedings of the 4th ACM workshop on Digital
  identity management}, 2008.

\bibitem{ma2009beyond}
J.~Ma, L.~K. Saul, S.~Savage, and G.~M. Voelker, ``{Beyond blacklists: learning
  to detect malicious web sites from suspicious URLs},'' in \emph{Proceedings
  of the 15th ACM SIGKDD international conference on Knowledge discovery and
  data mining}, 2009.

\bibitem{mao2017phishing}
J.~Mao, W.~Tian, P.~Li, T.~Wei, and Z.~Liang, ``{Phishing-Alarm: Robust and
  Efficient Phishing Detection via Page Component Similarity},'' \emph{IEEE
  Access}, vol.~5, pp. 17\,020--17\,030, 2017.

\bibitem{alam2020phishing}
M.~N. Alam, D.~Sarma, F.~F. Lima, I.~Saha, S.~Hossain \emph{et~al.},
  ``{Phishing attacks detection using machine learning approach},'' in
  \emph{{3rd International Conference on Smart Systems and Inventive Technology
  (ICSSIT), IEEE}}, 2020.

\bibitem{jain2019machine}
A.~K. Jain and B.~B. Gupta, ``{A machine learning based approach for phishing
  detection using hyperlinks information},'' \emph{Journal of Ambient
  Intelligence and Humanized Computing, Springer}, vol.~10, no.~5, pp.
  2015--2028, 2019.

\bibitem{abdelnabi2020visualphishnet}
S.~Abdelnabi, K.~Krombholz, and M.~Fritz, ``{Visualphishnet: Zero-day phishing
  website detection by visual similarity},'' in \emph{Proceedings of the 2020
  ACM SIGSAC conference on computer and communications security}, 2020.

\bibitem{phishtank}
C.~T.~I. Group, ``{PhishTank},'' 2006, \url{http://www.phishtank.com/developer}
  [Accessed: October 14th, 2023].

\bibitem{UCI1}
L.~M. Rami~Mohammad, ``{Phishing Websites, UC Irvine Machine Learning
  Repository},'' 2015,
  \url{https://archive.ics.uci.edu/ml/datasets/phishing+websites} [Accessed:
  October 14th, 2023].

\bibitem{mendeley}
S.~Ariyadasa, S.~Fernando, and S.~Fernando, ``{Mendeley Data, Phishing Websites
  Dataset},'' 2021, \url{https://data.mendeley.com/datasets/n96ncsr5g4}
  [Accessed: October 14th, 2023].

\bibitem{alexa_db}
Amazon, ``{Alexa Top Sites},'' \url{http://www.alexa.com/topsites}.

\bibitem{common_crawl}
E.~Gil, S.~Rich, and L.~Greg, ``{Common Crawl},''
  \url{https://commoncrawl.org/} [Accessed: October 14th, 2023].

\bibitem{tranco_db}
L.~P. Victor, V.~G. Tom, T.~Samaneh, K.~Maciej, and J.~Wouter, ``{Top Sites
  Ranking},'' \url{https://tranco-list.eu/} [Accessed: October 14th, 2023].

\bibitem{dmoz}
S.~Rick and T.~Bob, ``{Open Directory Project (ODP), DMOZ},'' 1998,
  \url{https://dmoz-odp.org/Computers/Artificial_Intelligence/Machine_Learning/Datasets/}
  [Accessed: October 14th, 2023].

\bibitem{curlie}
``{Curlie},'' 2017, \url{https://curlie.org/} [Accessed: October 14th, 2023].

\bibitem{openphish}
``{OpenPhish},'' 2014, \url{https://openphish.com/} [Accessed: October 14th,
  2023].

\bibitem{millersmiles}
M.~Bright, ``{MillerSmiles},'' 2003, \url{http://www.millersmiles.co.uk/}
  [Accessed: October 14th, 2023].

\bibitem{ebbu_2017}
Ebub, ``{Phishing Dataset},'' 2017,
  \url{https://github.com/ebubekirbbr/pdd/tree/master/input} [Accessed: October
  14th, 2023].

\bibitem{phishrepo}
S.~Ariyadasa, S.~Fernando, and S.~Fernando, ``{Mendeley Data,
  phishrepo-dataset},'' 2022,
  \url{https://data.mendeley.com/datasets/ttmmtsgbs8/4} [Accessed: October
  14th, 2023].

\bibitem{whois}
R.~Penman, ``{Python-whois},'' \url{https://pypi.org/project/python-whois/},
  2022, [Accessed: October 14th, 2023].

\bibitem{RDAP}
I.~E.~T. Force, ``{Registration Data Access Protocol},''
  \url{https://en.wikipedia.org/wiki/Registration_Data_Access_Protocol}, 2015,
  [Accessed: October 14th, 2023].

\bibitem{liu1998feature}
H.~Liu and H.~Motoda, \emph{{Feature extraction, construction and selection: A
  data mining perspective}}.\hskip 1em plus 0.5em minus 0.4em\relax Springer
  Science \& Business Media, 1998, vol. 453.

\bibitem{sarker2020context}
I.~H. Sarker, H.~Alqahtani, F.~Alsolami, A.~I. Khan, Y.~B. Abushark, and M.~K.
  Siddiqui, ``{Context pre-modeling: an empirical analysis for classification
  based user-centric context-aware predictive modeling},'' \emph{Journal Of Big
  Data, SpringerOpen}, vol.~7, no.~1, pp. 1--23, 2020.

\bibitem{pedregosa2011scikit}
F.~Pedregosa, G.~Varoquaux, A.~Gramfort, V.~Michel, B.~Thirion, O.~Grisel,
  M.~Blondel, P.~Prettenhofer, R.~Weiss, V.~Dubourg \emph{et~al.},
  ``{Scikit-learn: Machine learning in Python},'' \emph{the Journal of machine
  Learning research, JMLR. org}, vol.~12, pp. 2825--2830, 2011.

\bibitem{michalak2006correlation}
K.~Michalak and H.~Kwa{\'s}nicka, ``{Correlation-based feature selection
  strategy in classification problems},'' \emph{International Journal of
  Applied Mathematics and Computer Science, Uniwersytet Zielonog{\'o}rski.
  Oficyna Wydawnicza}, vol.~16, no.~4, pp. 503--511, 2006.

\bibitem{dubois1990rough}
D.~Dubois and H.~Prade, ``{Rough fuzzy sets and fuzzy rough sets},''
  \emph{International Journal of General System, Taylor \& Francis}, vol.~17,
  no. 2-3, pp. 191--209, 1990.

\bibitem{jalilifard2021semantic}
A.~Jalilifard, V.~F. Carid{\'a}, A.~F. Mansano, R.~S. Cristo, and F.~P.~C.
  da~Fonseca, ``{Semantic sensitive TF-IDF to determine word relevance in
  documents},'' in \emph{Advances in Computing and Network Communications:
  Proceedings of CoCoNet, Springer}, 2021, vol.~2, pp. 327--337.

\bibitem{abdi2010principal}
H.~Abdi and L.~J. Williams, ``{Principal Component Analysis},'' \emph{Wiley
  interdisciplinary reviews: computational statistics, Wiley Online Library},
  vol.~2, no.~4, pp. 433--459, 2010.

\bibitem{breiman2001random}
L.~Breiman, ``{Random forests},'' \emph{Machine learning, Springer}, vol.~45,
  pp. 5--32, 2001.

\bibitem{quinlan2014c4}
J.~R. Quinlan, \emph{{C4. 5: programs for machine learning}}.\hskip 1em plus
  0.5em minus 0.4em\relax Elsevier, 2014.

\bibitem{gordon1984classification}
A.~Gordon, L.~Breiman, J.~Friedman, R.~Olshen, and C.~J. Stone,
  ``{Classification and Regression Trees.}'' \emph{Biometrics, JSTOR}, vol.~40,
  no.~3, p. 874, 1984.

\bibitem{bharadiya2023review}
J.~P. Bharadiya, ``{A Review of Bayesian Machine Learning Principles, Methods,
  and Applications},'' \emph{International Journal of Innovative Science and
  Research Technology}, vol.~8, no.~5, pp. 2033--2038, 2023.

\bibitem{goodfellow2016deep}
I.~Goodfellow, Y.~Bengio, and A.~Courville, \emph{{Deep learning}}.\hskip 1em
  plus 0.5em minus 0.4em\relax MIT press, 2016.

\bibitem{sharma2017era}
P.~Sharma and A.~Singh, ``{Era of deep neural networks: A review},'' in
  \emph{8th International Conference on Computing, Communication and Networking
  Technologies (ICCCNT), IEEE}, 2017, pp. 1--5.

\bibitem{lecun1998gradient}
Y.~LeCun, L.~Bottou, Y.~Bengio, and P.~Haffner, ``{Gradient-based learning
  applied to document recognition},'' \emph{Proceedings of the IEEE}, vol.~86,
  no.~11, pp. 2278--2324, 1998.

\bibitem{huang2012semi}
W.~Huang, Z.~M. Tan, Z.~Lin, G.-B. Huang, J.~Zhou, C.-K. Chui, Y.~Su, and
  S.~Chang, ``{A semi-automatic approach to the segmentation of liver
  parenchyma from 3D CT images with Extreme Learning Machine},'' in
  \emph{Annual international conference of the IEEE engineering in medicine and
  biology society, IEEE}, 2012, pp. 3752--3755.

\bibitem{keerthi2001improvements}
S.~S. Keerthi, S.~K. Shevade, C.~Bhattacharyya, and K.~R.~K. Murthy,
  ``{Improvements to Platt's SMO algorithm for SVM classifier design},''
  \emph{Neural computation, MIT Press One Rogers Street, Cambridge, MA
  02142-1209, USA journals-info}, vol.~13, no.~3, pp. 637--649, 2001.

\bibitem{huang2015sequential}
X.~Huang, L.~Shi, and J.~A. Suykens, ``{Sequential minimal optimization for SVM
  with pinball loss},'' \emph{Neurocomputing}, vol. 149, pp. 1596--1603, 2015.

\bibitem{gupta2022pca}
I.~Gupta, V.~Sharma, S.~Kaur, and A.~K. Singh, ``{PCA-RF: an efficient
  Parkinson's disease prediction model based on random forest
  classification},'' \emph{arXiv preprint arXiv:2203.11287}, 2022.

\bibitem{witten2002data}
I.~H. Witten and E.~Frank, ``{Data mining: practical machine learning tools and
  techniques with Java implementations},'' \emph{ACM Sigmod Record, New York,
  NY, USA}, vol.~31, no.~1, pp. 76--77, 2002.

\bibitem{holte1993very}
R.~C. Holte, ``{Very simple classification rules perform well on most commonly
  used datasets},'' \emph{Machine learning, Springer}, vol.~11, pp. 63--90,
  1993.

\bibitem{pouriyeh2017comprehensive}
S.~Pouriyeh, S.~Vahid, G.~Sannino, G.~De~Pietro, H.~Arabnia, and J.~Gutierrez,
  ``{A comprehensive investigation and comparison of machine learning
  techniques in the domain of heart disease},'' in \emph{IEEE symposium on
  computers and communications (ISCC)}, 2017, pp. 204--207.

\bibitem{cessie1992ridge}
S.~l. Cessie and J.~V. Houwelingen, ``{Ridge estimators in logistic
  regression},'' \emph{Journal of the Royal Statistical Society Series C:
  Applied Statistics, Oxford University Press}, vol.~41, no.~1, pp. 191--201,
  1992.

\bibitem{john2013estimating}
G.~H. John and P.~Langley, ``{Estimating continuous distributions in Bayesian
  classifiers},'' \emph{arXiv preprint arXiv:1302.4964}, 2013.

\bibitem{aha1991instance}
D.~W. Aha, D.~Kibler, and M.~K. Albert, ``{Instance-based learning
  algorithms},'' \emph{Machine learning, Springer}, vol.~6, pp. 37--66, 1991.

\bibitem{freitas2002data}
A.~A. Freitas, \emph{{Data mining and knowledge discovery with evolutionary
  algorithms}}.\hskip 1em plus 0.5em minus 0.4em\relax Springer Science \&
  Business Media, 2002.

\bibitem{halgavs2020catching}
L.~Halga{\v{s}}, I.~Agrafiotis, and J.~R. Nurse, ``{Catching the Phish:
  Detecting Phishing Attacks using Recurrent Neural Networks (RNNs)},'' in
  \emph{Information Security Applications: 20th International Conference, WISA
  2019, Jeju Island, South Korea, August 21--24, Revised Selected Papers 20,
  Springer}, 2020.

\bibitem{o2015introduction}
K.~O'Shea and R.~Nash, ``{An Introduction to Convolutional Neural Networks},''
  \emph{arXiv preprint arXiv:1511.08458}, 2015.

\bibitem{van2020review}
G.~Van~Houdt, C.~Mosquera, and G.~N{\'a}poles, ``{A review on the long
  short-term memory model},'' \emph{Artificial Intelligence Review, Springer},
  vol.~53, pp. 5929--5955, 2020.

\bibitem{hinton2006fast}
G.~E. Hinton, S.~Osindero, and Y.-W. Teh, ``{A fast learning algorithm for deep
  belief nets},'' \emph{Neural computation}, vol.~18, no.~7, pp. 1527--1554,
  2006.

\bibitem{hua2015deep}
Y.~Hua, J.~Guo, and H.~Zhao, ``{Deep belief networks and deep learning},'' in
  \emph{Proceedings of International Conference on Intelligent Computing and
  Internet of Things, IEEE}, 2015.

\bibitem{rao2019detection}
R.~S. Rao and A.~R. Pais, ``{Detection of phishing websites using an efficient
  feature-based machine learning framework},'' \emph{{Neural Computing and
  Applications, Springer}}, vol.~31, no.~8, pp. 3851--3873, 2019.

\bibitem{somesha2020efficient}
M.~Somesha, A.~R. Pais, R.~S. Rao, and V.~S. Rathour, ``{Efficient deep
  learning techniques for the detection of phishing websites},''
  \emph{{S{\=a}dhan{\=a}, Springer}}, vol.~45, no.~1, pp. 1--18, 2020.

\bibitem{yi2018web}
P.~Yi, Y.~Guan, F.~Zou, Y.~Yao, W.~Wang, and T.~Zhu, ``{Web phishing detection
  using a deep learning framework},'' \emph{Wireless Communications and Mobile
  Computing, Hindawi}, vol. 2018, 2018.

\bibitem{medvet2008visual}
E.~Medvet, E.~Kirda, and C.~Kruegel, ``{Visual-similarity-based phishing
  detection},'' in \emph{{Proceedings of the 4th international conference on
  Security and privacy in communication networks, ACM}}, 2008.

\bibitem{fu2006detecting}
A.~Y. Fu, L.~Wenyin, and X.~Deng, ``{Detecting phishing web pages with visual
  similarity assessment based on earth mover's distance (EMD)},'' \emph{IEEE
  transactions on dependable and secure computing}, vol.~3, no.~4, pp.
  301--311, 2006.

\bibitem{afroz2011phishzoo}
S.~Afroz and R.~Greenstadt, ``{PhishZoo: Detecting Phishing Websites by Looking
  at Them},'' in \emph{IEEE fifth international conference on semantic
  computing}, 2011.

\bibitem{lin2021phishpedia}
Y.~Lin, R.~Liu, D.~M. Divakaran, J.~Y. Ng, Q.~Z. Chan, Y.~Lu, Y.~Si, F.~Zhang,
  and J.~S. Dong, ``{Phishpedia: A Hybrid Deep Learning Based Approach to
  Visually Identify Phishing Webpages},'' in \emph{USENIX Security Symposium},
  2021.

\bibitem{pan2006anomaly}
Y.~Pan and X.~Ding, ``{Anomaly based web phishing page detection},'' in
  \emph{2006 22nd Annual Computer Security Applications Conference (ACSAC'06),
  IEEE}, 2006, pp. 381--392.

\bibitem{garera2007framework}
S.~Garera, N.~Provos, M.~Chew, and A.~D. Rubin, ``{A framework for detection
  and measurement of phishing attacks},'' in \emph{Proceedings of the 2007 ACM
  workshop on Recurring malcode}, 2007.

\bibitem{zhang2017two}
W.~Zhang, Q.~Jiang, L.~Chen, and C.~Li, ``{Two-stage ELM for phishing Web pages
  detection using hybrid features},'' \emph{World Wide Web, Springer}, vol.~20,
  no.~4, pp. 797--813, 2017.

\bibitem{zhang2007cantina}
Y.~Zhang, J.~I. Hong, and L.~F. Cranor, ``{Cantina: a content-based approach to
  detecting phishing web sites},'' in \emph{Proceedings of the 16th
  international conference on World Wide Web, ACM}, 2007.

\bibitem{xiang2011cantina+}
G.~Xiang, J.~Hong, C.~P. Rose, and L.~Cranor, ``{Cantina+ a feature-rich
  machine learning framework for detecting phishing web sites},'' \emph{ACM
  Transactions on Information and System Security (TISSEC)}, vol.~14, no.~2,
  pp. 1--28, 2011.

\bibitem{hara2009visual}
M.~Hara, A.~Yamada, and Y.~Miyake, ``{Visual similarity-based phishing
  detection without victim site information},'' in \emph{IEEE Symposium on
  Computational Intelligence in Cyber Security}, 2009.

\bibitem{le2018urlnet}
H.~Le, Q.~Pham, D.~Sahoo, and S.~C. Hoi, ``{URLNet: Learning a URL
  representation with deep learning for malicious URL detection},'' \emph{arXiv
  preprint arXiv:1802.03162}, 2018.

\bibitem{kumar2020phishing}
J.~Kumar, A.~Santhanavijayan, B.~Janet, B.~Rajendran, and B.~Bindhumadhava,
  ``{Phishing website classification and detection using machine learning},''
  in \emph{{International Conference on Computer Communication and Informatics
  (ICCCI), IEEE}}, 2020.

\bibitem{sharma2019phishalert}
B.~Sharma and P.~Singh, ``{PhishAlert: An Efficient Phishing URL Detection via
  Hybrid Methodology},'' \emph{International Journal of Innovative Technology
  and Exploring Engineering (IJITEE)}, vol.~8, 2019.

\bibitem{jain2018phish}
A.~K. Jain and B.~Gupta, ``{PHISH-SAFE: URL features-based phishing detection
  system using machine learning},'' in \emph{Cyber Security}.\hskip 1em plus
  0.5em minus 0.4em\relax Springer, 2018, pp. 467--474.

\bibitem{abdelhamid2017phishing}
N.~Abdelhamid, F.~Thabtah, and H.~Abdel-jaber, ``{Phishing detection: A recent
  intelligent machine learning comparison based on models content and
  features},'' in \emph{{International conference on Intelligence and Security
  Informatics (ISI), IEEE}}, 2017, pp. 72--77.

\bibitem{sonmez2018phishing}
Y.~S{\"o}nmez, T.~Tuncer, H.~G{\"o}kal, and E.~Avc{\i}, ``{Phishing web sites
  features classification based on extreme learning machine},'' in \emph{{6th
  International Symposium on Digital Forensic and Security (ISDFS), IEEE}},
  2018, pp. 1--5.

\bibitem{verma2015character}
R.~Verma and K.~Dyer, ``{On the character of phishing URLs: Accurate and robust
  statistical learning classifiers},'' in \emph{{Proceedings of the 5th ACM
  Conference on Data and Application Security and Privacy}}, 2015.

\bibitem{rao2020catchphish}
R.~S. Rao, T.~Vaishnavi, and A.~R. Pais, ``{CatchPhish: detection of phishing
  websites by inspecting URLs},'' \emph{{Journal of Ambient Intelligence and
  Humanized Computing, Springer}}, vol.~11, no.~2, pp. 813--825, 2020.

\bibitem{shirazi2018kn0w}
H.~Shirazi, B.~Bezawada, and I.~Ray, ``{``Kn0w Thy Doma1n Name" Unbiased
  Phishing Detection Using Domain Name Based Features},'' in \emph{{Proceedings
  of the 23nd ACM on symposium on access control models and technologies}},
  2018.

\bibitem{sahingoz2019machine}
O.~K. Sahingoz, E.~Buber, O.~Demir, and B.~Diri, ``{Machine learning based
  phishing detection from URLs},'' \emph{Expert Systems with Applications,
  Elsevier}, vol. 117, pp. 345--357, 2019.

\bibitem{jeeva2016intelligent}
S.~C. Jeeva and E.~B. Rajsingh, ``{Intelligent phishing url detection using
  association rule mining},'' \emph{Human-centric Computing and Information
  Sciences, Springer}, vol.~6, no.~1, pp. 1--19, 2016.

\bibitem{blum2010lexical}
A.~Blum, B.~Wardman, T.~Solorio, and G.~Warner, ``{Lexical feature based
  phishing URL detection using online learning},'' in \emph{Proceedings of the
  3rd ACM Workshop on Artificial Intelligence and Security}, 2010.

\bibitem{torroledo2018hunting}
I.~Torroledo, L.~D. Camacho, and A.~C. Bahnsen, ``{Hunting malicious TLS
  certificates with deep neural networks},'' in \emph{{Proceedings of the 11th
  ACM workshop on Artificial Intelligence and Security}}, 2018.

\bibitem{wang2008light}
Y.~Wang, R.~Agrawal, and B.-Y. Choi, ``{Light weight anti-phishing with user
  whitelisting in a web browser},'' in \emph{IEEE region 5 conference}, 2008.

\bibitem{teraguchi2004client}
N.~Teraguchi and J.~C. Mitchell, ``{Client-side defense against web-based
  identity theft},'' \emph{Computer Science Department, Stanford University},
  2004.

\bibitem{GSB}
G.~S. Browsing, ``{Making the world’s information safely accessible},''
  \url{https://safebrowsing.google.com/} [Accessed: October 14th, 2023].

\bibitem{netcraft}
R.~Woodley, ``{Globally trusted defense against cybercrime, Netcraft},'' 2022,
  \url{https://www.netcraft.com/} [Accessed: October 14th, 2023].

\bibitem{avast}
I.~Belcic, ``{The Essential Guide to Phishing: How it Works and How to Defend
  Against it},'' \url{https://www.avast.com/c-phishing} [Accessed: October
  14th, 2023].

\bibitem{quickheal}
Q.~H.~T. Security, ``{Phishing Protection -- Product Documentation},''
  \url{https://docs.quickheal.com/docs/qhts/protection/phishing-protection/}
  [Accessed: October 14th, 2023].

\bibitem{mcafee}
L.~McAfee, ``{Next-level confidence for identity, privacy, and device
  protection},''
  \url{https://www.mcafee.com/en-in/antivirus/mcafee-livesafe.html} [Accessed:
  October 14th, 2023].

\bibitem{vaderetro}
V.~Secure, ``{Anti-Phishing Tools and Information: IsItPhishing Threat
  Detection},'' \url{https://isitphishing.org/} [Accessed: October 14th, 2023].

\bibitem{abuse.ch}
Abuse, ``{SSL Blacklist},'' \url{https://sslbl.abuse.ch/} [Accessed: October
  14th, 2023].

\bibitem{censys.io}
``{Censys},'' \url{https://censys.io/}[Accessed: October 14th, 2023].

\bibitem{master_dataset}
J.~Lilo, ``{Detecting Malicious URL Machine Learning},'' 2018,
  \url{https://github.com/rlilojr/Detecting-Malicious-URL-Machine-Learning/blob/master/dataset.csv}[Accessed:
  October 14th, 2023].

\bibitem{majestic_reports}
``{The Majestic Million},'' 2004,
  \url{https://majestic.com/reports/majestic-million} [Accessed: October 14th,
  2023].

\bibitem{yandex}
``{Yadex, Protection from Internet Fraud},'' 1997,
  \url{https://yandex.com/dev/safebrowsing/doc/quickstart/concepts/about.html}
  [Accessed: October 14th, 2023].

\bibitem{marchal2014phishstorm}
S.~Marchal, J.~Fran{\c{c}}ois, R.~State, and T.~Engel, ``{PhishStorm: Detecting
  phishing with streaming analytics},'' \emph{IEEE Transactions on Network and
  Service Management}, vol.~11, no.~4, pp. 458--471, 2014.

\bibitem{stuffgate}
``{Stuff Gate Website Outlook},'' 2012,
  \url{https://stuffgate.com.websiteoutlook.com/} [Accessed: October 14th,
  2023].

\bibitem{payment_service_providers}
``{List of online payment service providers},'' 2018,
  \url{https://research.omicsgroup.org/index.php/List_of_online_payment_service_providers}[Accessed:
  October 14th, 2023].

\bibitem{aburrous2010intelligent}
M.~Aburrous, M.~A. Hossain, K.~Dahal, and F.~Thabtah, ``{Intelligent phishing
  detection system for e-banking using fuzzy data mining},'' \emph{Expert
  systems with applications, Elsevier}, vol.~37, no.~12, pp. 7913--7921, 2010.

\bibitem{whittaker2010large}
C.~Whittaker, B.~Ryner, and M.~Nazif, ``{Large-scale automatic classification
  of phishing pages},'' \emph{{Google Research}}, 2010.

\bibitem{he2011efficient}
M.~He, S.-J. Horng, P.~Fan, M.~K. Khan, R.-S. Run, J.-L. Lai, R.-J. Chen, and
  A.~Sutanto, ``{An efficient phishing webpage detector},'' \emph{Expert
  systems with applications, Elsevier}, vol.~38, no.~10, pp. 12\,018--12\,027,
  2011.

\bibitem{el2017detection}
E.-S.~M. El-Alfy, ``{Detection of phishing websites based on probabilistic
  neural networks and K-medoids clustering},'' \emph{The Computer Journal,
  Oxford University Press}, vol.~60, no.~12, pp. 1745--1759, 2017.

\bibitem{montazer2015detection}
G.~A. Montazer and S.~ArabYarmohammadi, ``{Detection of phishing attacks in
  Iranian e-banking using a fuzzy--rough hybrid system},'' \emph{Applied Soft
  Computing, Elsevier}, vol.~35, pp. 482--492, 2015.

\bibitem{jacobs1995fast}
C.~E. Jacobs, A.~Finkelstein, and D.~H. Salesin, ``{Fast multiresolution image
  querying},'' in \emph{{SIGGRAPH'95: Proceedings of the 22nd annual conference
  on Computer graphics and interactive techniques, ACM}}, 1995.

\bibitem{corbetta2014eyes}
J.~Corbetta, L.~Invernizzi, C.~Kruegel, and G.~Vigna, ``{Eyes of a human, eyes
  of a program: Leveraging different views of the web for analysis and
  detection},'' in \emph{International Workshop on Recent Advances in Intrusion
  Detection, Springer}, 2014, pp. 130--149.

\bibitem{rubner2000earth}
Y.~Rubner, C.~Tomasi, and L.~J. Guibas, ``{The earth mover's distance as a
  metric for image retrieval},'' \emph{International journal of computer
  vision, Springer}, vol.~40, pp. 99--121, 2000.

\bibitem{burger2022scale}
W.~Burger and M.~J. Burge, ``{Scale-invariant feature transform (SIFT)},'' in
  \emph{Digital Image Processing: An Algorithmic Introduction, Springer}, 2022.

\bibitem{chicco2021siamese}
D.~Chicco, ``{Siamese neural networks: An overview},'' \emph{Artificial neural
  networks, Springer}, pp. 73--94, 2021.

\bibitem{menze2011oblique}
B.~H. Menze, B.~M. Kelm, D.~N. Splitthoff, U.~Koethe, and F.~A. Hamprecht,
  ``{On oblique random forests},'' in \emph{Machine Learning and Knowledge
  Discovery in Databases: European Conference, ECML PKDD 2011, Athens, Greece,
  Proceedings, Part II 22, Springer}, 2011.

\bibitem{amiri2014machine}
I.~S. Amiri, O.~A. Akanbi, and E.~Fazeldehkordi, \emph{{A machine-learning
  approach to phishing detection and defense}}.\hskip 1em plus 0.5em minus
  0.4em\relax Syngress, Elsevier, 2014.

\bibitem{bahnsen2017classifying}
A.~C. Bahnsen, E.~C. Bohorquez, S.~Villegas, J.~Vargas, and F.~A. Gonz{\'a}lez,
  ``{Classifying phishing URLs using recurrent neural networks},'' in
  \emph{{APWG symposium on electronic crime research (eCrime), IEEE}}, 2017.

\bibitem{rao2019jail}
R.~S. Rao and A.~R. Pais, ``{Jail-Phish: An improved search engine based
  phishing detection system},'' \emph{Computers \& Security, Elsevier},
  vol.~83, pp. 246--267, 2019.

\bibitem{alkawaz2021comprehensive}
M.~H. Alkawaz, S.~J. Steven, A.~I. Hajamydeen, and R.~Ramli, ``{A Comprehensive
  Survey on Identification and Analysis of Phishing Website based on Machine
  Learning Methods},'' in \emph{11th IEEE Symposium on Computer Applications \&
  Industrial Electronics (ISCAIE)}, 2021.

\bibitem{abu2007comparison}
S.~Abu-Nimeh, D.~Nappa, X.~Wang, and S.~Nair, ``{A comparison of machine
  learning techniques for phishing detection},'' in \emph{Proceedings of the
  anti-phishing working groups 2nd annual eCrime researchers summit, ACM},
  2007.

\bibitem{zabihimayvan2019fuzzy}
M.~Zabihimayvan and D.~Doran, ``{Fuzzy rough set feature selection to enhance
  phishing attack detection},'' in \emph{International Conference on Fuzzy
  Systems, IEEE}, 2019.

\bibitem{jain2018two}
A.~K. Jain and B.~B. Gupta, ``{Two-level authentication approach to protect
  from phishing attacks in real time},'' \emph{Journal of Ambient Intelligence
  and Humanized Computing, Springer}, vol.~9, pp. 1783--1796, 2018.

\bibitem{corona2017deltaphish}
I.~Corona, B.~Biggio, M.~Contini, L.~Piras, R.~Corda, M.~Mereu, G.~Mureddu,
  D.~Ariu, and F.~Roli, ``{Deltaphish: Detecting Phishing Webpages in
  Compromised Websites},'' in \emph{22nd European Symposium on Research in
  Computer Security, Oslo, Norway, Proceedings, Part I 22, Computer
  Security--ESORICS, Springer}, 2017.

\bibitem{chhabra2011phi}
S.~Chhabra, A.~Aggarwal, F.~Benevenuto, and P.~Kumaraguru, ``{Phi. sh/\$ ocial:
  the phishing landscape through short urls},'' in \emph{Proceedings of the 8th
  Annual Collaboration, Electronic messaging, Anti-Abuse and Spam Conference
  (CEAS), ACM}, 2011.

\bibitem{tang2021survey}
L.~Tang and Q.~H. Mahmoud, ``{A survey of machine learning-based solutions for
  phishing website detection},'' \emph{Machine Learning and Knowledge
  Extraction, MDPI}, vol.~3, no.~3, pp. 672--694, 2021.

\bibitem{basnet2014learning}
R.~B. Basnet, A.~H. Sung, and Q.~Liu, ``{Learning to detect phishing URLs},''
  \emph{International Journal of Research in Engineering and Technology
  (IJRET)}, vol.~3, no.~6, pp. 11--24, 2014.

\bibitem{pylongurl}
R.~B. Basnet, ``{PyLongURL - Python library for longurl.org},''
  \url{https://storage.googleapis.com/google-code-archive-downloads/v2/code.google.com/pylongurl/PyLongURL.py},
  2010, [Accessed: October 14th, 2023].

\bibitem{sameen2020phishhaven}
M.~Sameen, K.~Han, and S.~O. Hwang, ``{PhishHaven—An efficient real-time AI
  phishing URLs detection system},'' \emph{IEEE Access}, vol.~8, pp.
  83\,425--83\,443, 2020.

\bibitem{adversarialAttackDefinition}
G.~Ian, P.~Nicolas, H.~Sandy, D.~Yan, A.~Pieter, and C.~Jack, ``{Attacking
  machine learning with adversarial examples: OpenAI},''
  \url{https://openai.com/research/attacking-machine-learning-with-adversarial-examples},
  2017, [Accessed: October 14th, 2023].

\bibitem{das2019sok}
A.~Das, S.~Baki, A.~El~Aassal, R.~Verma, and A.~Dunbar, ``{SoK: a comprehensive
  reexamination of phishing research from the security perspective},''
  \emph{IEEE Communications Surveys \& Tutorials}, vol.~22, no.~1, pp.
  671--708, 2019.

\bibitem{dalvi2004adversarial}
N.~Dalvi, P.~Domingos, Mausam, S.~Sanghai, and D.~Verma, ``{Adversarial
  classification},'' in \emph{Proceedings of the tenth ACM SIGKDD international
  conference on Knowledge discovery and data mining}, 2004.

\bibitem{shirazi2019adversarial}
H.~Shirazi, B.~Bezawada, I.~Ray, and C.~Anderson, ``{Adversarial sampling
  attacks against phishing detection},'' in \emph{Data and Applications
  Security and Privacy XXXIII: 33rd Annual IFIP WG 11.3 Conference, DBSec 2019,
  Charleston, SC, USA, July 15--17, 2019, Proceedings 33, Springer}, 2019.

\bibitem{openai_chatgpt}
OpenAI, ``{Introducing ChatGPT},'' \url{https://openai.com/blog/chatgpt}, 2022,
  [Accessed: October 14th, 2023].

\bibitem{linda_mini_tool}
M.~Linda, ``{ChatGPT, This Content May Violate Our Content Policy},''
  \url{https://www.minitool.com/news/chatgpt-this-content-may-violate-our-content-policy.html},
  2023, [Accessed: October 14th, 2023].

\bibitem{roy2023generating}
S.~S. Roy, K.~V. Naragam, and S.~Nilizadeh, ``{Generating Phishing Attacks
  using ChatGPT},'' \emph{arXiv preprint arXiv:2305.05133}, 2023.

\bibitem{alabdan2020phishing}
R.~Alabdan, ``{Phishing attacks survey: Types, vectors, and technical
  approaches},'' \emph{Future internet, MDPI}, vol.~12, no.~10, p. 168, 2020.

\bibitem{kang2010captcha}
L.~Kang and J.~Xiang, ``{CAPTCHA phishing: A practical attack on human
  interaction proofing},'' in \emph{Information Security and Cryptology: 5th
  International Conference, Inscrypt 2009, Beijing, China, Springer}, 2010.

\bibitem{alnajjar2016trustqr}
A.~Y. Alnajjar, S.~Manickam, M.~Anbar, S.~Al-Saleem, and O.~Elejla, ``{TrustQR:
  A new technique for the detection of phishing attacks on QR code},''
  \emph{Advanced Science Letters, American Scientific Publishers}, vol.~22,
  no.~10, pp. 2905--2909, 2016.

\bibitem{liang2016cracking}
B.~Liang, M.~Su, W.~You, W.~Shi, and G.~Yang, ``{Cracking classifiers for
  evasion: A case study on the google's phishing pages filter},'' in
  \emph{Proceedings of the 25th International Conference on World Wide Web,
  ACM}, 2016.

\bibitem{vargas2016knowing}
J.~Vargas, A.~C. Bahnsen, S.~Villegas, and D.~Ingevaldson, ``{Knowing your
  enemies: Leveraging data analysis to expose phishing patterns against a major
  US financial institution},'' in \emph{APWG Symposium on Electronic Crime
  Research (eCrime), IEEE}, 2016.

\bibitem{bahnsen2018deepphish}
A.~C. Bahnsen, I.~Torroledo, L.~D. Camacho, and S.~Villegas, ``{DeepPhish:
  Simulating Malicious AI},'' in \emph{{APWG symposium on electronic crime
  research (eCrime)}}, 2018.

\bibitem{koide2023detecting}
T.~Koide, N.~Fukushi, H.~Nakano, and D.~Chiba, ``{Detecting Phishing Sites
  Using ChatGPT},'' \emph{arXiv preprint arXiv:2306.05816}, 2023.

\bibitem{forbes}
T.~F. Sunny~Chu, Scott~Filiault, ``{Forbes Insights, Analytics: Don't Forget
  The Human Element},''
  \url{https://images.forbes.com/forbesinsights/StudyPDFs/EY-DataandAnalyticsImpactIndex-REPORT.pdf}
  [Accessed: October 14th, 2023].

\bibitem{interbrand}
E.~Walkom, ``{Interbrand: Best Global Brands 2022},''
  \url{https://interbrand.com/newsroom/interbrand-launches-best-global-brands-2022/}
  [Accessed: October 14th, 2023].

\bibitem{fortune}
M.~Stankova, ``{FORTUNE 500 List Of Companies 2022 And Their Domain Name
  Choices},''
  \url{https://smartbranding.com/fortune-500-list-of-companies-2022-and-their-domain-name-choices/}
  [Accessed: October 14th, 2023].

\bibitem{bloomberg}
E.~Chang, ``{Bloomberg, Top Headlines: Google Seeks Top-Level Domains},''
  \url{https://www.bloomberg.com/news/articles/2012-05-31/top-headlines-google-seeks-top-level-domains}
  [Accessed: October 14th, 2023].

\bibitem{brandz}
Kantar, ``{BrandZ Top 100 Most Valuable Global Brands},''
  \url{https://www.rankingthebrands.com/The-Brand-Rankings.aspx?rankingID=6}
  [Accessed: October 14th, 2023].

\bibitem{kumaraguru2010teaching}
P.~Kumaraguru, S.~Sheng, A.~Acquisti, L.~F. Cranor, and J.~Hong, ``{Teaching
  Johnny not to fall for phish},'' \emph{ACM Transactions on Internet
  Technology (TOIT)}, vol.~10, no.~2, pp. 1--31, 2010.

\bibitem{vijayalakshmi2020web}
M.~Vijayalakshmi, S.~Mercy~Shalinie, M.~H. Yang, and R.~M. U, ``{Web phishing
  detection techniques: a survey on the state-of-the-art, taxonomy and future
  directions},'' \emph{Iet Networks, Wiley Online Library}, vol.~9, no.~5, pp.
  235--246, 2020.

\bibitem{patil2019methodical}
S.~Patil and S.~Dhage, ``{A methodical overview on phishing detection along
  with an organized way to construct an anti-phishing framework},'' in
  \emph{5th International Conference on Advanced Computing \& Communication
  Systems (ICACCS), IEEE}, 2019.

\bibitem{do2022deep}
N.~Q. Do, A.~Selamat, O.~Krejcar, E.~Herrera-Viedma, and H.~Fujita, ``{Deep
  Learning for Phishing Detection: Taxonomy, Current Challenges and Future
  Directions},'' \emph{IEEE Access}, 2022.

\bibitem{hannousse2021towards}
A.~Hannousse and S.~Yahiouche, ``{Towards benchmark datasets for machine
  learning based website phishing detection: An experimental study},''
  \emph{Engineering Applications of Artificial Intelligence, Elsevier}, vol.
  104, p. 104347, 2021.

\bibitem{zieni2023phishing}
R.~Zieni, L.~Massari, and M.~C. Calzarossa, ``{Phishing or not phishing? A
  survey on the detection of phishing websites},'' \emph{IEEE Access}, vol.~11,
  pp. 18\,499--18\,519, 2023.

\bibitem{catal2022applications}
C.~Catal, G.~Giray, B.~Tekinerdogan, S.~Kumar, and S.~Shukla, ``{Applications
  of deep learning for phishing detection: a systematic literature review},''
  \emph{Knowledge and Information Systems, Springer}, vol.~64, no.~6, pp.
  1457--1500, 2022.

\bibitem{sahoo2017malicious}
D.~Sahoo, C.~Liu, and S.~C. Hoi, ``{Malicious URL detection using machine
  learning: A survey},'' \emph{arXiv preprint arXiv:1701.07179}, 2017.

\bibitem{jain2022survey}
A.~K. Jain and B.~Gupta, ``{A survey of phishing attack techniques, defence
  mechanisms and open research challenges},'' \emph{Enterprise Information
  Systems, Taylor \& Francis}, vol.~16, no.~4, pp. 527--565, 2022.

\bibitem{singh2020phishing}
C.~Singh \emph{et~al.}, ``{Phishing website detection based on machine
  learning: A survey},'' in \emph{2020 6th International Conference on Advanced
  Computing and Communication Systems (ICACCS), IEEE}, 2020.

\bibitem{basit2021comprehensive}
A.~Basit, M.~Zafar, X.~Liu, A.~R. Javed, Z.~Jalil, and K.~Kifayat, ``{A
  comprehensive survey of AI-enabled phishing attacks detection techniques},''
  \emph{{Telecommunication Systems, Springer}}, vol.~76, no.~1, pp. 139--154,
  2021.

\bibitem{prachit2020survey}
R.~Prachit, V.~Harshal, and R.~Shete, ``{A Survey of Phishing Website Detection
  Systems},'' \emph{International Research Journal of Engineering and
  Technology (IRJET)}, vol.~7, pp. 1145--1148, 2020.

\bibitem{abdillah2022phishing}
R.~Abdillah, Z.~Shukur, M.~Mohd, and T.~M.~Z. Murah, ``{Phishing classification
  techniques: A systematic literature review},'' \emph{IEEE Access}, vol.~10,
  pp. 41\,574--41\,591, 2022.

\bibitem{safi2023systematic}
A.~Safi and S.~Singh, ``{A systematic literature review on phishing website
  detection techniques},'' \emph{Journal of King Saud University-Computer and
  Information Sciences, Elsevier}, 2023.

\bibitem{rastogi2021survey}
M.~Rastogi, A.~Chhetri, D.~K. Singh \emph{et~al.}, ``{Survey on detection and
  prevention of phishing websites using machine learning},'' in
  \emph{International Conference on Advance Computing and Innovative
  Technologies in Engineering (ICACITE), IEEE}, 2021.

\bibitem{aung2019survey}
E.~S. Aung, C.~T. Zan, and H.~Yamana, ``{A survey of URL-based phishing
  detection},'' in \emph{DEIM Forum}, 2019.

\bibitem{khonji2013phishing}
M.~Khonji, Y.~Iraqi, and A.~Jones, ``{Phishing detection: a literature
  survey},'' \emph{IEEE Communications Surveys \& Tutorials}, vol.~15, no.~4,
  pp. 2091--2121, 2013.

\bibitem{zuhair2016feature}
H.~Zuhair, A.~Selamat, and M.~Salleh, ``{Feature selection for phishing
  detection: a review of research},'' \emph{International Journal of
  Intelligent Systems Technologies and Applications, Inderscience Publishers
  (IEL)}, vol.~15, no.~2, pp. 147--162, 2016.

\end{thebibliography}

\end{document}